\documentclass[nofootinbib,aps,11pt,amsmath,amssymb,a4paper,prd]{revtex4-1}

\usepackage[english,american]{babel}
\usepackage[utf8]{inputenc}
\usepackage[T1]{fontenc}

\usepackage{graphicx}
\usepackage{color}

\usepackage{slashed}

\usepackage{units}

\usepackage{hyperref}
\hypersetup{
   pdfnewwindow=true,    
   colorlinks=true,      
   linkcolor=black,       
   citecolor=blue,       
   filecolor=black,       
   urlcolor=black         
}

\begin{document}

\title{\Large {\bf{Theory for Baryon Number and Dark Matter at the LHC}}}
\author{Michael \surname{Duerr}}
\email{michael.duerr@mpi-hd.mpg.de}
\author{Pavel \surname{Fileviez P\'erez}}
\email{fileviez@mpi-hd.mpg.de}
\affiliation{\\ \small{
Max-Planck-Institut f\"ur Kernphysik, \\ Saupfercheckweg 1, 69117 Heidelberg, Germany}
}

\begin{abstract}
We investigate the possibility to test a simple theory for spontaneous baryon number violation at the Large Hadron Collider. 
In this context the baryon number is a local gauge symmetry spontaneously broken at the low scale through the Brout--Englert--Higgs mechanism. This theory predicts the existence of a leptophobic neutral gauge boson and a fermionic dark matter candidate with baryon number. We study the gauge boson and Higgs decays, and explore the connection between collider signatures and constraints coming 
from dark matter experiments. We point out an upper bound on the symmetry breaking scale using the relic density constraints, which tells us that this model can be tested or ruled out at current or future collider experiments.
\end{abstract}

\maketitle

\tableofcontents

\newpage

\section{Introduction}
The nature of dark matter~(DM) in the Universe is one of the big open questions in science. Even though no direct evidence for the particle nature of this 
non-luminous but gravitationally interacting form of matter has been found yet, dark matter is embraced by the particle physics community as one of the 
most pressing issues demanding an extension of the well-established Standard Model~(SM) of particle physics. A particle candidate for dark matter has to 
be stable or at least long-lived on cosmological time scales. Therefore, it is crucial to understand dynamically the stability of the dark matter.
Symmetries provide a nice way to guarantee the stability of a dark matter candidate, but often such a symmetry is only imposed by hand in a particular 
dark matter model, without following from an underlying principle. 

There is a long history of attempts to consistently gauge the accidental global symmetries of baryon and lepton numbers~\cite{Pais:1973mi,Rajpoot:1987yg,Foot:1989ts,Carone:1995pu,Georgi:1996ei,FileviezPerez:2010gw,Dulaney:2010dj,FileviezPerez:2011pt,Duerr-PFP-Wise} of the SM Lagrangian. Recently, simple realistic versions of such theories for local baryon and lepton numbers have been discussed~\cite{FileviezPerez:2010gw,Dulaney:2010dj,FileviezPerez:2011pt,Duerr-PFP-Wise,Arnold:2013qja,PFP-Ohmer-Patel,Duerr:2013lka,Hiren}. Anomaly cancellation requires the introduction of additional fermions, and new fields charged under $U(1)_B$ and $U(1)_L$---we refer to them as ``lepto-baryons''---provide a simple solution~\cite{Duerr-PFP-Wise, PFP-Ohmer-Patel}. The dark matter stability is automatic in these theories since after symmetry breaking there is a remnant $\mathcal{Z}_2$ symmetry in the new sector that forbids the decay of the lightest new particle with fractional baryon number. 

These theories have a rich phenomenology and it is worthwhile to study them in more detail. One of the most interesting predictions is that when local baryon number is broken at the low scale the proton is stable and there is no need to postulate the existence of a large desert. The cosmological aspects of these models have been investigated 
in Refs.~\cite{Dulaney:2010dj,Hiren}, where it has been shown that even though baryon number is broken at the low scale, a non-zero baryon asymmetry can be generated in agreement 
with the experiment.  

In this article, we discuss phenomenological aspects of a minimal theory for spontaneous baryon number violation~\cite{Duerr-PFP-Wise,Duerr:2013lka} in detail. 
Our main goal is to understand the testability of this model. Therefore, we discuss the properties of the new leptophobic gauge boson related to baryon number, as well as properties of the additional physical Higgs field. We present their decay properties as well as possible production channels at the LHC. We investigate the predictions for dark matter experiments using 
the relevant collider constraints. Finally, we discuss the possibility to find an upper bound on the symmetry breaking scale using the relic density constraints. This upper bound 
tells us that this model could be tested or ruled out at current or future colliders. 

This article is organized as follows. In Section~\ref{sec:localbaryonnumber} we review simple theories for local baryon number. In Section~\ref{sec:lhc} we discuss phenomenological aspects of one of these theories at the LHC in detail. We study the dark matter sector of this theory in Section~\ref{sec:dm}, and finally we summarize our main results 
in Section~\ref{sec:summary}. 
\section{Theories for Local Baryon Number}
\label{sec:localbaryonnumber}
Recently, simple theories where the baryon and lepton numbers are local gauge symmetries have been proposed and investigated in some detail~\cite{FileviezPerez:2010gw,Dulaney:2010dj,FileviezPerez:2011pt,Duerr-PFP-Wise,PFP-Ohmer-Patel,Duerr:2013lka,Hiren}. These theories are based on the gauge group 
\begin{equation}\label{eq:GBL}
 G_{BL} = SU(3) \otimes SU(2) \otimes U(1)_Y \otimes U(1)_B \otimes U(1)_L.
\end{equation}
There are two simple realistic versions of these theories: one can add new vector-like fields with baryon and lepton numbers to cancel all the anomalies~\cite{Duerr-PFP-Wise}, see the particle content beyond the Standard Model in Table~\ref{tab:ModelI}, or have new fields with Majorana masses after symmetry breaking~\cite{PFP-Ohmer-Patel}. 
After symmetry breaking, both models have the same extra degrees of freedom: only eight, even if the number of representations is different. Since the new fields carry both baryon and lepton number, we refer to them as ``lepto-baryons''.  

\begin{table}[t]
   \caption{Fermion content beyond the Standard Model and the corresponding quantum numbers of the lepto-baryons in the model~\cite{Duerr-PFP-Wise} to be studied in detail in this article.
The right-handed neutrinos $\nu^{\alpha}_R$ are present in the model ($\alpha=1,2,3$ is the family index).}
\label{tab:ModelI}
\begin{tabular}{cccccc}
\hline\hline
~~Fields ~~  &~~ $SU(3)$ ~~ &~~ $SU(2)$~~ &~~ $U(1)_Y$~~ &~~ $U(1)_B$~~ &~~ $U(1)_L$ ~~ \\
\hline 
   $\nu^{\alpha}_R$   & {\bf 1}  & {\bf 1} & 0 & 0 & 1 \\
   $\Psi_L$ & {\bf 1} & {\bf 2} & - $\frac{1}{2}$ & $B_1$ & $L_1$ \\ 
   $\Psi_R$ & {\bf 1} & {\bf 2} & - $\frac{1}{2}$ & $B_2$ & $L_2$ \\ 
   $\eta_R$ & {\bf 1} & {\bf 1} & - $1$ & $B_1$ & $L_1$ \\ 
   $\eta_L$ & {\bf 1} & {\bf 1} & - ${1}$ & $B_2$ & $L_2$ \\ 
   $\chi_R$ & {\bf 1} & {\bf 1} & 0 & $B_1$ & $L_1$ \\   
   $\chi_L$ & {\bf 1} & {\bf 1} & 0 & $B_2$ & $L_2$ \\ 
\hline \hline
\end{tabular}
\end{table}

In this article we discuss phenomenological and cosmological aspects of the model with the particle content given in Table~\ref{tab:ModelI}~\cite{Duerr-PFP-Wise}. 
We focus only on the sector with local baryon number such that the relevant gauge group is given by
\begin{equation}
 G_{B} = SU(3) \otimes SU(2) \otimes U(1)_Y \otimes U(1)_B,
 \end{equation}
which is obtained after the lepton number is broken at a high scale. 

The Lagrangian of the model is given by
\begin{equation}
 \mathcal{L} = \mathcal{L}_\text{SM} + \mathcal{L}_B,
\end{equation} 
where $\mathcal{L}_\text{SM}$ is the Lagrangian for the SM fields. Since all quarks have baryon number, their kinetic term is modified, taking into account an additional coupling to the new gauge boson related to baryon number. The new part of the Lagrangian, $\mathcal{L}_B$, is given by 
\begin{equation}
\label{eq:LB}
 \mathcal{L}_B = - \frac{1}{4} B_{\mu \nu}^B B^{\mu \nu, B}  - \frac{\epsilon_{B}}{2} B_{\mu \nu}^B B^{\mu \nu} +  \mathcal{L}_f +   \mathcal{L}_{S_B},
\end{equation} 
where
$B_{\mu \nu}= \partial_\mu B_\nu - \partial_\nu B_\mu$ is the $U(1)_Y$ field strength tensor and $B_{\mu \nu}^B= \partial_\mu B_\nu^B - \partial_\nu B_\mu^B$ is the $U(1)_B$ field strength tensor. 
The coupling $\epsilon_B$ encapsulates the kinetic mixing between hypercharge and baryon number.  
The term $\mathcal{L}_f$ contains the couplings of the new fermions,
\begin{align}
\label{eq:Lf}
 \mathcal{L}_f   & =  i \overline{\Psi}_L \slashed{D} \Psi_L + i \overline{\Psi}_R \slashed{D} \Psi_R + i \overline{\eta}_L \slashed{D} \eta_L +  i \overline{\eta}_R \slashed{D} \eta_R + i \overline{\chi}_L \slashed{D} \chi_L + i \overline{\chi}_R \slashed{D} \chi_R \nonumber \\
& \quad -  Y_1 \overline{\Psi}_L H \eta_R - Y_2 \overline{\Psi}_L \tilde{H} \chi_R  
 - Y_3 \overline{\Psi}_R H \eta_L - Y_4 \overline{\Psi}_R \tilde{H} \chi_L \nonumber \\
& \quad -  \lambda_{\Psi} \overline{\Psi}_L \Psi_R S_{B} - \lambda_{\eta} \overline{\eta}_R \eta_L S_{B} 
 - \lambda_{\chi} \overline{\chi}_R \chi_L S_{B} \ + \ {\rm h.c.},
\end{align}
where the Standard Model Higgs field $H$ and the additional scalar boson $S_B$ transform as
\begin{equation}
 H \sim (\mathbf{1},\mathbf{2},1/2,0) \qquad \text{and}  \qquad S_{B} \sim (\mathbf{1},\mathbf{1},0,B_1-B_2),
\end{equation}
and $\tilde{H} = i \sigma_2 H^\ast$. The term $\mathcal{L}_{S_B}$ is defined as 
\begin{equation}
\mathcal{L}_{S_B} =   \left( D_\mu S_B\right)^\dagger D^\mu S_B - m_B^2 S_B^\dagger S_B  - \lambda_B (S_B^\dagger S_B)^2  - \lambda_{HB} (H^\dagger H)(S_B^\dagger S_B) .
\end{equation} 

The baryon numbers $B_1$ and $B_2$ of the new fermions are constrained by the conditions of anomaly cancellation (see the detailed discussion in Refs.~\cite{Duerr-PFP-Wise,PFP-Ohmer-Patel}), and one finds that all relevant anomalies are cancelled for any choice of $B_1$ and $B_2$ which satisfy the condition
\begin{equation}
\label{eq:conditionB}
 B_1 - B_2 = -3.
\end{equation}
We will therefore treat $B \equiv B_1 + B_2$ as a free parameter in the rest of the article. Notice that couplings that generate Majorana masses for the SM singlet fields $\chi_L$ and $\chi_R$ after symmetry breaking such as $\chi_L \chi_L S_B$ and $\chi_R \chi_R S_B^\dagger$ would be allowed for $B_1 = - B_2$. We will discuss only the Dirac case, sticking to $B_1 \neq -B_2$ in the remainder of the article.

The condition in Eq.~\eqref{eq:conditionB} and the need to generate vector-like masses for the new fermions unambiguously fix the baryon number of the new Higgs boson $S_B$, such that it transforms as
\begin{equation}
S_B \sim  (\mathbf{1},\mathbf{1},0,-3).
\end{equation}
Therefore, once $S_B$ obtains a vacuum expectation value breaking local baryon number, we will only have $|\Delta B|=3$ interactions and proton decay never occurs.
This is a key result which tells us that the great desert is not needed to suppress proton decay and the cutoff of the theory can be low.

A few comments regarding the $Z$--$Z_B$ mixing are in order. Because the SM Higgs doublet does not carry baryon number and we break $U(1)_B$ by a SM singlet, there is no mass mixing between the two gauge bosons and the mixing angle $\xi$ is induced by $\epsilon_B$ only~\cite{Babu:1997st},
\begin{equation}\label{eq:mixing-angle}
\tan 2 \xi = \frac{-2 \hat{M}_Z^2 \hat{s}_W \epsilon_B \sqrt{1 - \epsilon_B^2}}{\hat{M}_{Z_B}^2-\hat{M}_Z^2 (1-\epsilon_B^2) + \hat{M}_Z^2 \hat{s}_W^2 \epsilon_B^2},
\end{equation}
where $\hat{M}_Z$ is the mass of the $Z$ boson, $\hat{M}_{Z_B}$ is the mass of the $Z_B$, and $\hat{s}_W$ is the sine of the Weinberg angle. Here, the hats refer to the fact that the parameters are coming from the original mixed basis, where the gauge fields have non-diagonal kinetic terms. See Ref.~\cite{Babu:1997st} for a general discussion of kinetic mixing. Precision measurements constrain the mixing $\xi$ to be small, see Ref.~\cite{Erler:2009jh} for current constraints, which are of the order $10^{-3}$.

Even if the mixing vanishes at tree level, it will arise at loop level because we have fermions in the theory that carry both baryon number and non-zero hypercharge. However, for small $\epsilon_B$ we may approximate 
\begin{equation}
 \tan 2 \xi \simeq - \frac{2 \hat{M}_Z^2 \hat{s}_W \epsilon_B}{\hat{M}_{Z_B}^2},
\end{equation}
and it is easy to see that a (small) loop-induced $\epsilon_B$ is even further suppressed by the large mass of the $Z_B$. See Appendix~\ref{app:kineticmixing} for more details, where the loop-induced kinetic mixing is calculated and shown to be within the experimental limits easily. We will therefore neglect the kinetic mixing $\epsilon_B$ in the remainder of the article. 

Let us also comment on the spectrum of the theory. After spontaneous breaking of baryon number, we have four charged and four neutral new fermions in the theory. These fermions carry baryon number, see Table~\ref{tab:ModelI}, and do not couple directly to SM quarks and leptons. Therefore, no new sources of flavor violation in the SM quark and lepton sectors are introduced. 
There is a remnant $\mathcal{Z}_2$ symmetry after breaking of the local $U(1)_B$ under which the new fermions are odd whereas all  other fields are even. Thus, the lightest new fermion with fractional baryon number is automatically stable and---if neutral---a DM candidate. Notice that this stability is a direct consequence of symmetry breaking and does not have to be put into the theory by hand.
We will assume that the mixing between the $SU(2)$ doublets and singlets in the new sector is small, i.e., we assume small Yukawa couplings $Y_i$ ($i=1,\ldots,4$) in Eq.~\eqref{eq:Lf}, and take the SM singlet-like Dirac field $\chi = \chi_L + \chi_R$ to be our DM candidate. The rest of the new fermions is assumed to be heavy. 
We will discuss bounds on the mass of the second physical Higgs field in Sec.~\ref{sec:higgs}. 
\section{Phenomenological Aspects at the LHC}
\label{sec:lhc}
Our main goal is to understand the testability of the model. Therefore, one needs to identify the properties of the decays of the leptophobic gauge boson $Z_B$ and the new Higgs field $h_2$ and understand the connection to dark matter. We will focus on the most optimistic scenario where
$M_{Z_B}, M_{h_2} > 2 M_\chi$. In this case both the leptophobic gauge boson and the new physical Higgs can decay into dark matter and one can realize the test of this model at collider experiments. Therefore, one can distinguish this theory from other scenarios where the baryon number is a local symmetry but there is no dark matter candidate.
\subsection{Leptophobic Gauge Boson Decays}
The model predicts the existence of a new neutral gauge boson associated with the local baryon number. The interactions of this gauge boson that are relevant for our discussion are given by
\begin{equation}
\mathcal{L}_B \supset - g_B \bar{\chi} \left( B_1 P_L \ + \ B_2 P_R \right)   \gamma^\mu  \chi \  Z_\mu^B + \frac{1}{2} M_{Z_B}^2 Z_\mu^B Z^{B, \mu} - \frac{1}{3} g_B \sum _i \bar{q}_i \gamma^\mu q_i Z_\mu^B,
\end{equation}    
where we have neglected the (small) kinetic mixing $\epsilon_B$ between the two Abelian symmetries for simplicity, such that $Z_\mu^B = B_\mu^B$. Then, the mass of the leptophobic gauge boson is given by 
\begin{equation}
 M_{Z_B}=3 g_B v_B,
\end{equation}
where $v_B$ is the vacuum expectation value of the $S_B$ boson. 
Here $P_L=(1-\gamma^5)/2$ and $P_R=(1+\gamma^5)/2$ are the usual left- and right-handed projection operators, while $B_1$ and $B_2$ are the baryon numbers of the new fermions, see Table~\ref{tab:ModelI} for the assignment. 
 Motivated by the dark matter study in Section~\ref{sec:dm} we assume that the $Z_B$ gauge boson decays only into all the Standard Model quarks and into dark matter $\chi$.

\begin{figure}[t]
\includegraphics[width=0.49\linewidth]{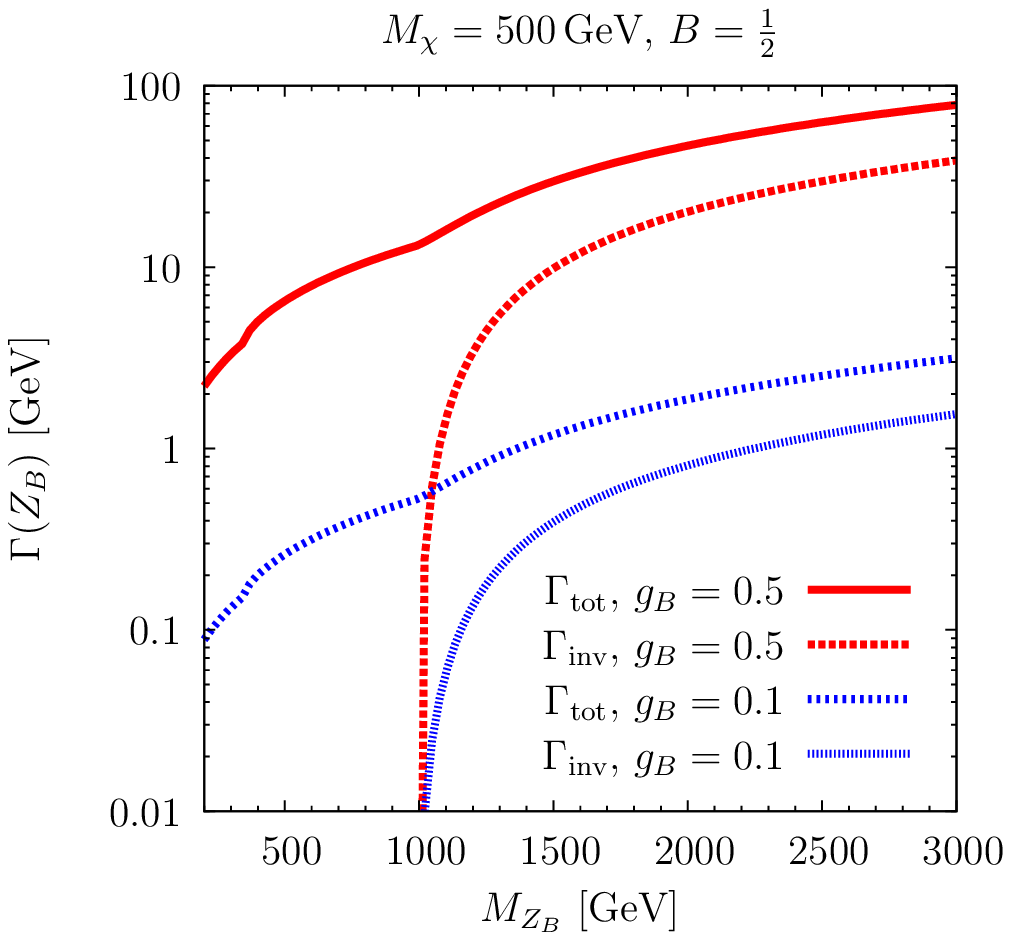}
\includegraphics[width=0.49\linewidth]{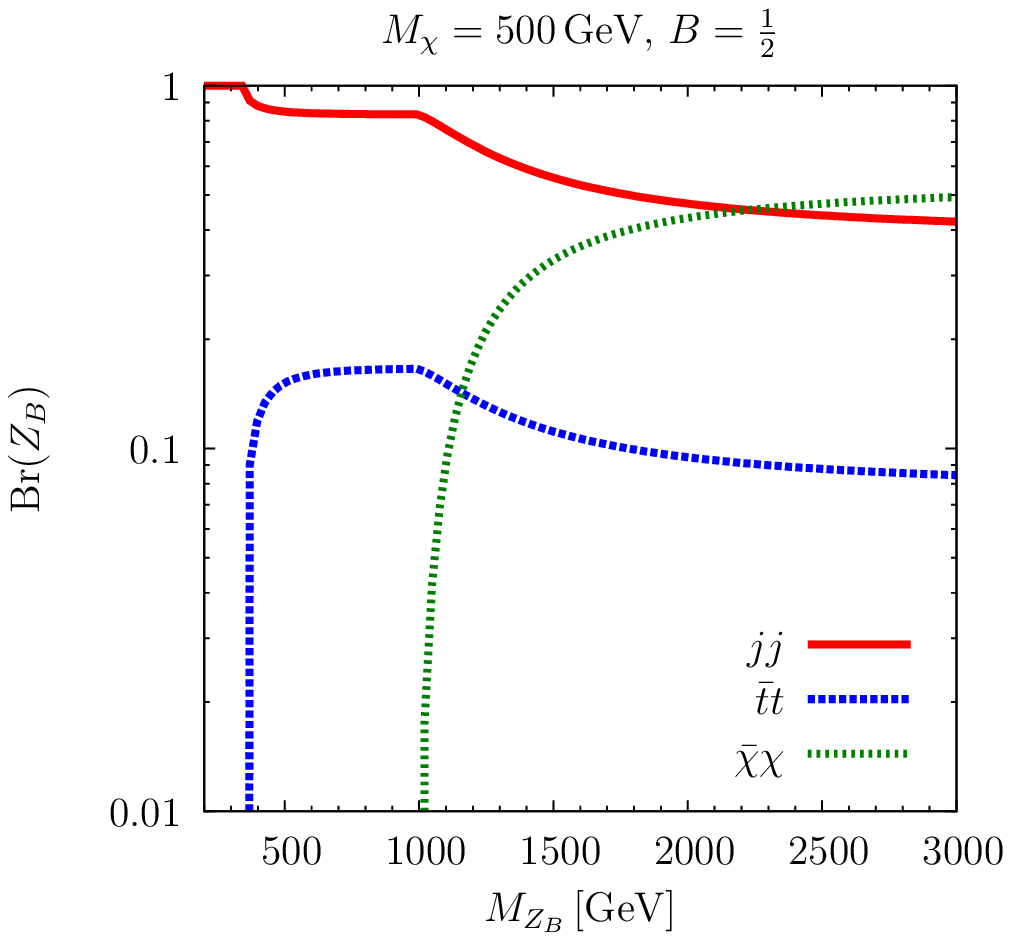}
 \caption{Decays of the leptophobic gauge boson $Z_B$. The left panel shows the total width $\Gamma_\text{tot}(Z_B)$ and the invisible width $\Gamma_\text{inv}(Z_B)$ for two values of the gauge coupling, $g_B = 0.1$ (in blue) and $g_B = 0.5$ (in red). The right panel shows the branching ratios of the $Z_B$ decays into jets (in red), top quark pairs (in blue), and dark matter (in green). Notice that the branching ratios are independent of the value of $g_B$. The plots are for a dark matter mass $M_\chi = \unit[500]{GeV}$ and $B \equiv B_1+ B_2=1/2$.}
 \label{fig:ZBprops}
\end{figure}

The $Z_B$ gauge boson can be produced at the LHC through its coupling to quarks. Its decay properties are given in Fig.~\ref{fig:ZBprops}, where the left panel shows the total width $\Gamma_\text{tot}(Z_B)$ and the invisible width $\Gamma_\text{inv}(Z_B)$ for two values of the gauge coupling,  $g_B = 0.1$ and $g_B = 0.5$, and the right panel shows the branching ratios to jets, top quark pairs, and dark matter. In both panels, the dark matter mass is set to $M_\chi = \unit[500]{GeV}$ and $B = 1/2$. 

For a given mass $M_{Z_B}$ of the leptophobic gauge boson, one can use the left panel of Fig.~\ref{fig:ZBprops} to read off the corresponding total width $\Gamma_\text{tot} (Z_B)$ for a particular gauge coupling $g_B$, such that a measurement of the total width could give us the value of the gauge coupling. For example, using $M_{Z_B} = \unit[1.5]{TeV}$, one obtains $\Gamma_\text{tot} (Z_B) = \unit[1.19]{GeV}$ for $g_B = 0.1$ and $\Gamma_\text{tot} (Z_B) = \unit[29.7]{GeV}$ for $g_B = 0.5$. Notice that the contribution of the invisible decay into dark matter to the total width above threshold could be large. It is important to have this invisible decay to test the model. Having only decays into the Standard Model quarks, it would be difficult to distinguish this model from other $Z^\prime$ models where there is no relation between baryon number and the dark matter sector. 

The right panel of Fig.~\ref{fig:ZBprops} shows the $Z_B$ branching ratios, which are independent of the choice of $g_B$. In most of the parameter space, the branching ratio into top quark pairs is about $\text{Br}(Z_B \to \bar{t}t) \approx 0.1$. For large $M_{Z_B}$, the decay into dark matter is possible and may even  become dominant, for the given parameters up to $\text{Br} ( Z_B \to \bar{\chi} \chi )  \approx 0.5$ around $M_{Z_B} = \unit[3]{TeV}$.

Using the decay of $Z_B$ into two top quarks, the ATLAS collaboration has set bounds on this type of gauge bosons~\cite{ATLAS}. The relevant process is the decay into two top quarks with mass $M_t$,
$$pp \ \to \ Z_B^* \ \to \ \bar{t} t.$$ 
The hadronic production cross section for this process is given by
\begin{equation}
\sigma (p p \to Z_B^* \to  \bar{t} t ) (s) = \int_{\tau_0}^1 d \tau \frac{d {\cal L}^{pp}_{q \bar{q}}}{d \tau}  \ \sigma (q \bar{q} \to Z_B^* \to  \bar{t} t ) (\hat{s}).
 \end{equation}
It can be computed using the cross section at the partonic level
\begin{equation}
\sigma (\bar{q} q \to Z_B^* \to  \bar{t} t ) (\hat{s})= \frac{g_B^4 \sqrt{\hat{s} - 4 M_t^2}}{972 \pi \sqrt{\hat{s}}} \frac{\left(  2 M_t^2 + \hat{s} \right)}{\left[ (\hat{s} - M_{Z_B}^2)^2 + M_{Z_B}^2 \Gamma_{Z_B}^2 \right]},
\end{equation}
together with the MSTW 2008 parton distribution functions~\cite{Martin:2009iq} giving the parton luminosities via 
\begin{equation}
\label{eq:partonluminosities}
\frac{d {\cal L}^{pp}_{q \bar{q}}}{d \tau} = \int_{\tau}^1 \frac{d x}{x} \left[ f_{q/p} \left(x, \mu \right)  f_{\bar{q}/p} \left(\frac{\tau}{x}, \mu \right) 
+ f_{q/p} \left(\frac{\tau}{x}, \mu \right)  f_{\bar{q}/p} \left(x, \mu \right)\right].
\end{equation}
Here, $\tau=\hat{s}/s$, $\hat{s}$ is the partonic center-of-mass energy squared, $s$ is the hadronic center-of-mass energy squared, $\tau_0=4 M_t^2/s$ is the production threshold, and $\mu$ is the factorization scale. We use the abbreviation $\Gamma_{Z_B} = \Gamma_\text{tot} (Z_B)$.

\begin{figure}[t]
\includegraphics[width=0.49\linewidth]{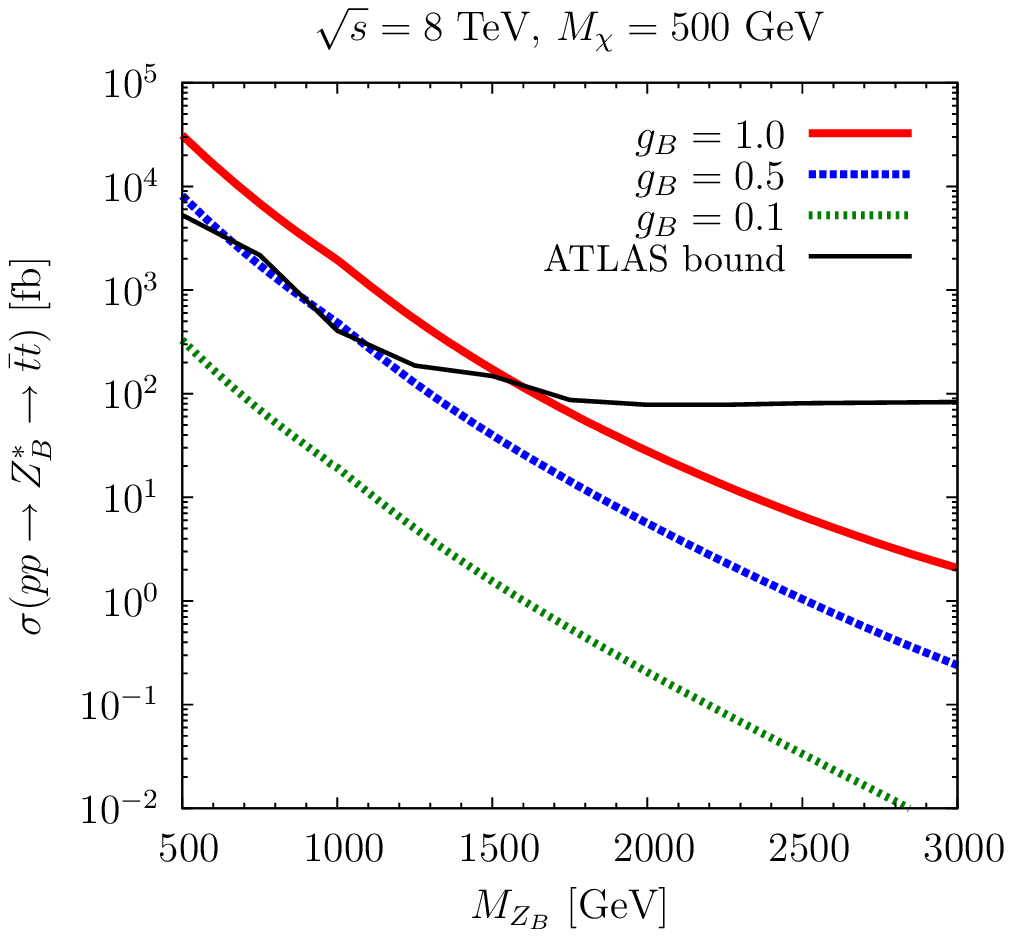}
\includegraphics[width=0.49\linewidth]{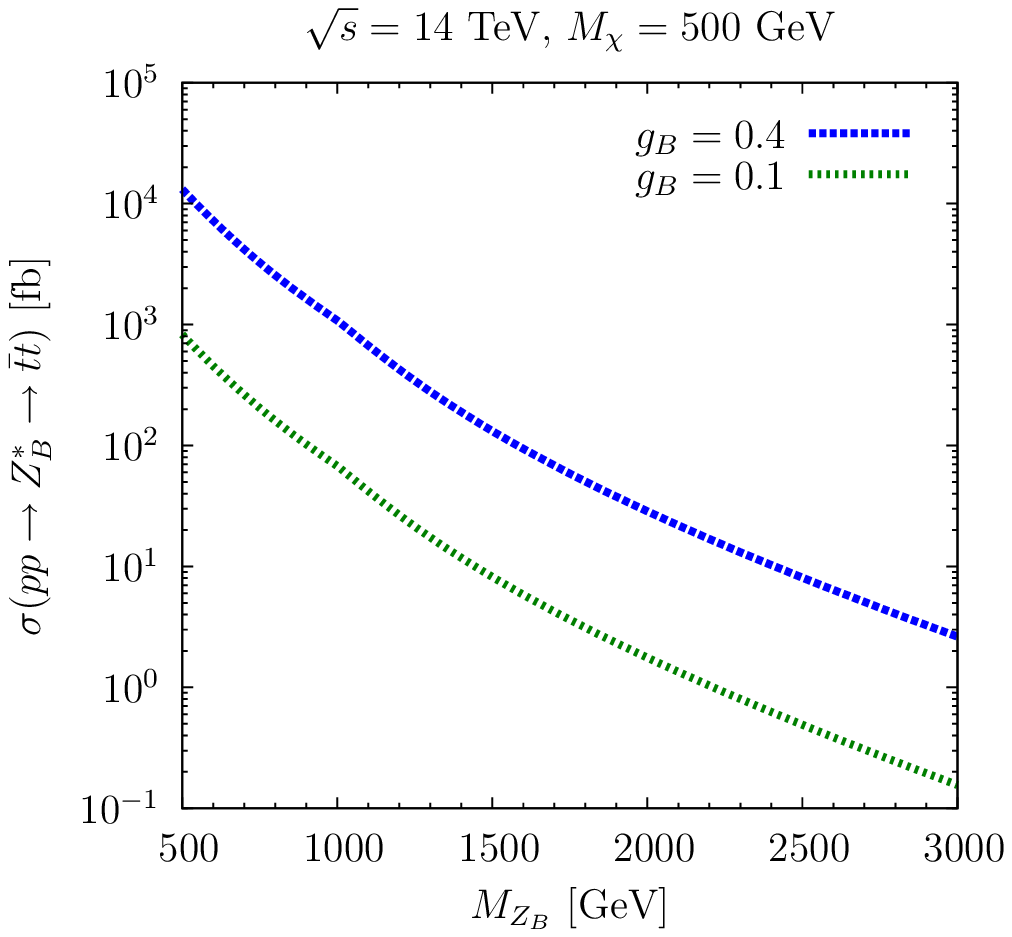}
 \caption{Decay of the leptophobic gauge boson $Z_B$ into two top quarks. The left panel shows the experimental bounds from the ATLAS collaboration~\cite{ATLAS} (solid black) and the theoretical predictions for different values of the gauge couplings ($g_B=1$ in red, $g_B=0.5$ in blue, and $g_B=0.1$ in green) when $\sqrt{s}=\unit[8]{TeV}$. The right panel shows the cross section for different values of the gauge couplings ($g_B=0.4$ in blue and $g_B=0.1$ in green) when $\sqrt{s}=\unit[14]{TeV}$. Notice that we set (as before) $M_\chi = \unit[500]{GeV}$, which has impact on the cross section through the decay width $\Gamma_{Z_B}$.}
 \label{fig:ZBbounds}
\end{figure}

Using these equations, we show the numerical results for the cross section in Fig.~\ref{fig:ZBbounds} (left panel) for different values of the gauge coupling, $g_B = 0.1$, $0.5$, and $1.0$, for a center-of-mass energy of $\sqrt{s} = \unit[8]{TeV}$. The black curve is the experimental upper bound from the ATLAS collaboration~\cite{ATLAS}. We use a $K$-factor of $K= 1.1$ to account for next-to-leading order QCD effects in the $Z_B$ production and decay into $\bar{t}t$~\cite{Caola:2012rs}. Notice that this is smaller than the value $K=1.3$ typically adopted by ATLAS~\cite{ATLAS}, because a detailed calculation shows that additional negative contributions result in a reduction of the $K$-factor~\cite{Caola:2012rs}. Since the area above the black curve is ruled out by experiment, one can say that the gauge coupling must be smaller than 0.5 to be consistent with the experiment in most of the parameter space. This is the main result of this section. A value of $g_B = 0.5$ is ruled out or at least borderline for values $M_{Z_B} \leq \unit[1.1]{TeV}$, while for values $M_{Z_B} \geq \unit[1.6]{TeV}$, $g_B = 1$ is viable. We will use $g_B = 0.4$ as 
benchmark value in the rest of the article. Notice that the onset of decays into DM changes the width of the $Z_B$, in the plot this is the case at $M_{Z_B} = \unit[1]{TeV}$ because $M_\chi = \unit[500]{GeV}$, and the cross section is therefore modified. We will stick to $Z_B$'s with mass $M_{Z_B} \geq \unit[1]{TeV}$ in our discussion, for light $Z_B$'s with small gauge couplings see the discussion in Refs.~\cite{Dobrescu:2013cmh,An:2012ue,Dobrescu:2014fca}. We would like to mention that the bounds from mono-jet searches are very weak when the $Z_B$ gauge boson is heavy, see Ref.~\cite{Buchmueller:2014yoa} for details. 

In Fig.~\ref{fig:ZBbounds} (right panel) we show the numerical results for the center-of-mass energy $\sqrt{s}=\unit[14]{TeV}$ in order to understand the possibility to test this theory in the next run of the LHC. We can estimate the expected number of events by
\begin{equation}
 N(\bar{t} t )=\mathcal{L} \times \sigma (pp \to Z_B^\ast \to \bar{t} t).
\end{equation}
Assuming a luminosity of $\mathcal{L} = 30~\text{fb}^{-1}$, a gauge boson mass of $M_{Z_B} = \unit[1.5]{TeV}$ and a gauge coupling of $g_B=0.4$, one has $\sigma (pp \to Z_B^\ast \to \bar{t} t) = \unit[131.8]{fb}$, and thus one obtains $N(\bar{t} t )= 4.0 \times 10^3$.
Therefore, in this way we can probe most of the parameter space in this sector at the LHC even with ${\cal{L}}=30$ fb$^{-1}$. In order to distinguish between the signal 
and the large QCD background one needs to impose the standard cut on the invariant mass of two quarks, i.e., $M_{t\bar{t}} \approx M_{Z_B}$. See for example 
Ref.~\cite{ATLAS, CMS} for the reconstruction of these events. 

\subsection{Higgs Boson Decays and Production Mechanisms}\label{sec:higgs}
The Higgs sector of the model is composed of the SM Higgs $H$ and the $S_B$ boson breaking the local baryon number, which we write as
\begin{equation}
H^T= \left( 0 \  \  \frac{h(x) + v_0}{\sqrt{2}} \right) \quad  \text{and} \quad  S_B= \frac{(h_B(x) + v_B)}{\sqrt{2}} e^{i \sigma_B(x) / v_B}.
\end{equation}
Using the scalar potential at tree level,
\begin{equation}
V(H,S_B)=m_H^2 H^\dagger H + \lambda_H (H^\dagger H)^2 + m_B^2 S_B^\dagger S_B  + \lambda_B (S_B^\dagger S_B)^2  + \lambda_{HB} (H^\dagger H)(S_B^\dagger S_B),
\end{equation}
we find the minimization conditions
\begin{eqnarray}
v_0 \left(  m_H^2 + \lambda_H v_0^2 + \frac{1}{2} \lambda_{HB} \ v_B^2 \right)=0, \\
v_B \left( m_B^2 + \lambda_B v_B^2 + \frac{1}{2} \lambda_{HB} \ v_0^2 \right)=0.
\end{eqnarray}
Therefore, there are four possible vacua:
\begin{enumerate}

\item $v_0=0$ and $v_B=0$. This vacuum has zero energy, $V_\text{min}^{(1)}(0,0)=0$, and of course it is not phenomenologically viable. 

\item  $v_0 \neq 0$ and $v_B=0$. In this case we cannot generate vector-like masses for the new particles and the energy of the two degenerate vacua is given by
\begin{equation}
V_\text{min}^{(2)} (v_0,0)=-\frac{1}{8} \ M_h^2 \ v_0^2 \approx - \unit[1.2 \times 10^8]{GeV}^4,
\end{equation}
where $M_h=\unit[126]{GeV}$ is the SM Higgs mass and $v_0=\unit[246]{GeV}$ is the SM Higgs vacuum expectation value.  
\item $v_0=0$ and $v_B \neq 0$. The local baryon number symmetry is broken in this case and the energy of the minima is defined by
\begin{equation}
V_\text{min}^{(3)} (0,v_B)=-\frac{1}{4} \  \lambda_B \ v_B^4.
\end{equation}
As in the previous cases one cannot have a realistic scenario. 

\item $v_0 \neq 0$ and $v_B \neq 0$. This is the only realistic scenario and from the minimization conditions we find
\begin{eqnarray}
v_0^2=- 2 \frac{\left( 2 m_H^2 \lambda_B - m_B^2 \lambda_{HB} \right)}{4 \lambda_B \lambda_H - \lambda_{HB}^2 }, \ \ 
v_B^2=- 2 \frac{\left( 2 m_B^2 \lambda_H - m_H^2 \lambda_{HB} \right)}{4 \lambda_B \lambda_H - \lambda_{HB}^2 }. 
\end{eqnarray}

\end{enumerate}

We will of course stick to case 4. In order to have a potential bounded from below and a minimum we need to impose the conditions
\begin{equation}
\lambda_H > 0, \ \lambda_B > 0, \  {\rm and} \  \lambda_H \lambda_B - \frac{1}{4} \lambda_{HB}^2 >0.
\end{equation}
In this case the energy of the minima is given by
\begin{equation}
V_{\rm{min}} ^{(4)}(v_0,v_B)= -\frac{1}{4} \  \lambda_H \ v_0^4 - \frac{1}{4} \  \lambda_B \ v_B^4 -\frac{1}{4} \  \lambda_{HB} \ v_0^2 v_B^2.
\end{equation}
Therefore, using this expression we can set the condition to use the global minimum when $\lambda_{HB}$ is positive.
Now, we are ready to study the physical spectrum in this realistic scenario.
The mass matrix for the physical Higgses in the basis $(h, h_B)$ is given by
\begin{equation}
{\cal M}^2_0 = 
\left(
\begin{array}{ccc}
2 v_0^2 \lambda_H  &   v_0 v_B \lambda_{HB}   \\
 v_0 v_B \lambda_{HB}   &   2 v_B^2 \lambda_B  \\
\end{array}
\right),
\end{equation}
and the physical states are defined as
\begin{align}
 h_1 &= \cos \theta_B  \ h \ + \  \sin \theta_B \ h_B, \\
 h_2 & = - \sin \theta_B \ h \ + \  \cos \theta_B \ h_B,
\end{align}
where the mixing angle is
\begin{equation}
 \tan 2 \theta_B = \frac{v_0 v_B \lambda_{HB}}{v_0^2 \lambda_H - v_B^2 \lambda_B}.
\end{equation}

\begin{figure}[t]
\includegraphics[width=0.49\linewidth]{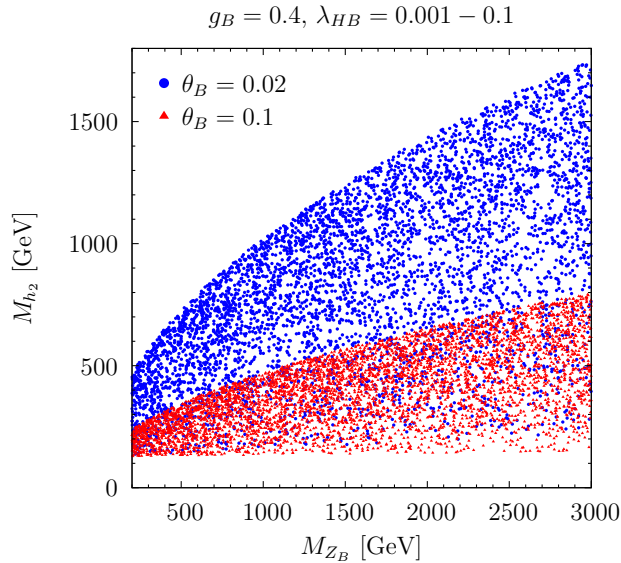}
 \caption{Mass of the heavy CP-even Higgs $h_2$ vs.\ the mass of the leptophobic gauge boson $Z_B$. We use $M_{h_1}=\unit[126]{GeV}$, $g_B = 0.4$ and vary $\lambda_{HB} \in [0.001,0.1]$. Blue dots are for $\theta_B = 0.02$, red triangles are for $\theta_B = 0.1$.}
 \label{fig:h2mass}
\end{figure}

\noindent The masses of the physical Higgs fields are
\begin{eqnarray}
M_{h_1}^2 &=& v_0^2 \lambda_H + v_B^2 \lambda_B -  |\csc 2 \theta_B| \ v_0 v_B \lambda_{HB} \approx \unit[126]{GeV}, \\
M_{h_2}^2 &=& v_0^2 \lambda_H + v_B^2 \lambda_B +  |\csc 2 \theta_B| \ v_0 v_B \lambda_{HB}.
\end{eqnarray}
These expressions are valid only when $\lambda_{HB} \neq 0$. Notice that when $\lambda_{HB}=0$ the two Higgses do not mix and we have the SM Higgs 
and the $h_B$ in the new sector. Here $v_0=\unit[246]{GeV}$, $v_B=M_{Z_B}/3 g_B$ and the Higgs masses are related via
\begin{equation}
M_{h_2}^2=M_{h_1}^2 \ + \  \frac{2}{3 g_B} | \csc 2 \theta_B | \ v_0 \ M_{Z_B} \lambda_{HB}. 
\end{equation} 
Using this expression, we show in Fig.~\ref{fig:h2mass} the numerical values for $M_{h_2}$ as a function of the input parameters. Notice that there is an upper limit on the mass of the $h_2$, depending of course on the value of the input parameters. In particular, for the used values, the $h_2$ is always lighter than the leptophobic gauge boson $Z_B$. 

After symmetry breaking both Higgses will have interactions with the SM fields, as well as self-interactions. 
The relevant interactions for our discussions are
\begin{align}
\mathcal{L} &\supset - \frac{M_f}{v_0} \cos \theta_B \ \bar{f} f h_1 \ + \  \frac{M_f}{v_0} \sin \theta_B \ \bar{f} f h_2 
+ \frac{2 M_W^2}{v_0} \cos \theta_B h_1 W_\mu W^\mu + \frac{M_Z^2}{v_0} \cos \theta_B h_1 Z_\mu Z^\mu \nonumber \\
&\quad - \frac{2 M_W^2}{v_0} \sin \theta_B h_2 W_\mu W^\mu - \frac{M_Z^2}{v_0} \sin \theta_B h_2 Z_\mu Z^\mu 
- \frac{M_\chi}{v_B} \sin \theta_B \ \bar{\chi} \chi h_1 \ - \  \frac{M_\chi}{v_B} \cos \theta_B \ \bar{\chi} \chi h_2 \nonumber \\
&\quad - \frac{1}{2} c_{112} \ h_1^2 h_2 + 6 g_B M_{Z_B} \cos\theta_B h_2 Z_\mu^B Z^{B,\mu},
\end{align}
where
\begin{align}
c_{112} &= -6 v_0 \lambda_H \cos^2 \theta_B \sin \theta_B \ + \  6 v_B  \lambda_B \cos \theta_B \sin^2 \theta_B \nonumber \\
 & \quad + \lambda_{HB} \left( v_B \cos^3 \theta_B + 2 v_0 \cos^2 \theta_B \sin \theta_B - 2 v_B \cos \theta_B \sin^2 \theta_B - v_0 \sin^3 \theta_B \right).
\end{align}
The parameters $\lambda_B$, $\lambda_H$ and $\lambda_{HB}$ can be written as functions of the other free parameters,
\begin{eqnarray}
\lambda_B&=& \frac{9 g_B^2}{M_{Z_B}^2}  \frac{\left( M_{h_1}^2 c_2 \ - \ M_{h_2}^2 c_1 \right)}{c_2^2 - c_1^2}, \\ 
\lambda_H&=& \frac{1}{v_0^2}  \frac{\left( M_{h_1}^2 c_1 \ - \ M_{h_2}^2 c_2 \right)}{c_1^2 - c_2^2}, \\
\lambda_{HB}&=& \frac{3 g_B \tan 2 \theta_B}{M_{Z_B} v_0}  \frac{\left( M_{h_1}^2 \ - \ M_{h_2}^2  \right)}{c_1-c_2},
\end{eqnarray}
where
\begin{eqnarray}
c_1&=&1 - |\csc 2 \theta_B| \tan 2 \theta_B, \ \ \ {\rm{and}} \  \
c_2=1 + |\csc 2 \theta_B| \tan 2 \theta_B.
\end{eqnarray}
Therefore, since $M_{h_1} \approx \unit[126]{GeV}$, this model has only six free parameters
\begin{displaymath}
g_B, \ M_{Z_B}, \  \theta_B, \ M_{h_2}, \ B, \ \text{and} \ M_\chi,
\end{displaymath}
and the complete discussion of its phenomenological and cosmological aspects can be done using these parameters.

The SM-like Higgs $h_1$ cannot decay into dark matter, $\chi$, or into two $h_2$ because the latter two are too heavy. Therefore, in this model the branching ratios of $h_1$ are the same as in the SM. However, the couplings of $h_1$ to fermions and to the gauge bosons contain the mixing angle via the factor $\cos \theta_B$, which is constrained by the experiments. For example, using the ratios~\cite{PDG}
\begin{eqnarray}
R_{\gamma \gamma}&=&\frac{\sigma (pp \to h_1) \times {\rm Br} (h_1 \to \gamma \gamma)}{\sigma (pp \to h)_\text{SM} \times {\rm Br} (h \to \gamma \gamma)_\text{SM}}= 1.58^{+0.27}_{-0.23}, \\
R_{WW}&=&\frac{\sigma (pp \to h_1) \times {\rm Br} (h_1 \to W W^*)}{\sigma (pp \to h)_\text{SM} \times {\rm Br} (h \to W W^*)_\text{SM}}= 0.87^{+0.24}_{-0.22}, \\
R_{ZZ}&=&\frac{\sigma (pp \to h_1) \times {\rm Br} (h_1 \to Z Z^*)}{\sigma (pp \to h)_\text{SM} \times {\rm Br} (h \to Z Z^*)_\text{SM}}= 1.11^{+0.34}_{-0.28},
\end{eqnarray}   
one can set a bound on the mixing angle. See Ref.~\cite{Englert:2014uua} for a discussion of the current constraints on Higgs couplings. Using the central value $R_{WW}=0.87$ we naively find $\theta_B = 0.37~(\sim \pi/10)$. 
We will discuss phenomenologically viable values for the mixing angle in more detail below.

\begin{figure}[t]
\includegraphics[width=0.5\linewidth]{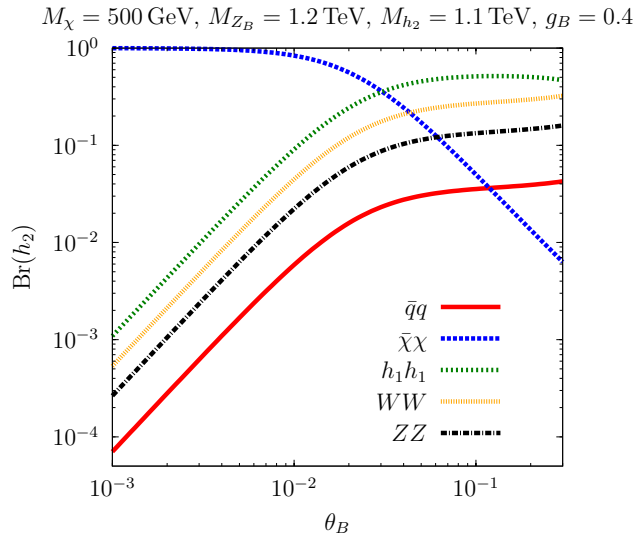}
 \caption{Branching ratios of $h_2$ as a function of the mixing angle $\theta_B$. Here we use $M_\chi=\unit[500]{GeV}$, $M_{Z_B}=\unit[1.2]{TeV}$, $M_{h_2}=\unit[1.1]{TeV}$, and $g_B=0.4$ as input parameters. The branching ratio to leptons is too small to be visible in the plot. Notice that the decay into two $Z_B$ is not allowed kinematically for the given choice of parameters.}
 \label{fig:h2BR}
\end{figure}

The heavy Higgs $h_2$ has interesting properties, because one can have the decays
\begin{equation*}
 h_2 \to \bar{q} q, \ \bar{e} e, \ WW, \ ZZ, \ h_1 h_1, \ \bar{\chi} \chi, \ Z_B Z_B.
\end{equation*}
In Fig.~\ref{fig:h2BR} we show the branching ratios as a function of the mixing angle $\theta_B$. For small mixing angles ($\theta_B \leq 0.03$), the decays into SM fields are suppressed, while the invisible decays into dark matter dominate. For large mixing angles ($\theta_B \gg 0.03$), the decay into dark matter is strongly suppressed, and the decay into the SM Higgs dominates over the decays into SM fermions and gauge bosons. The distinguishing feature of decays into dark matter motivates the use of a rather small mixing angle ($\theta_B = 0.02$) in the phenomenological survey of the model. 

 \begin{figure}
\includegraphics[width=0.435\linewidth]{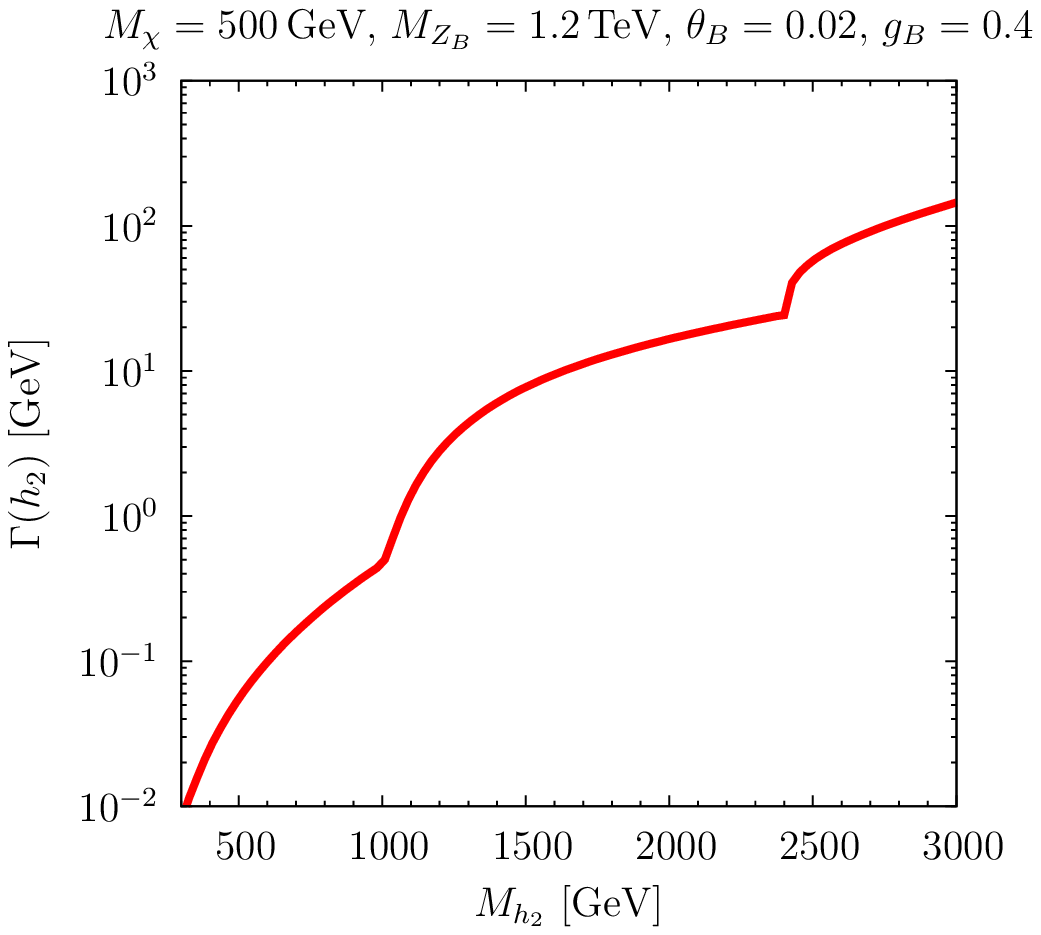}
\includegraphics[width=0.555\linewidth]{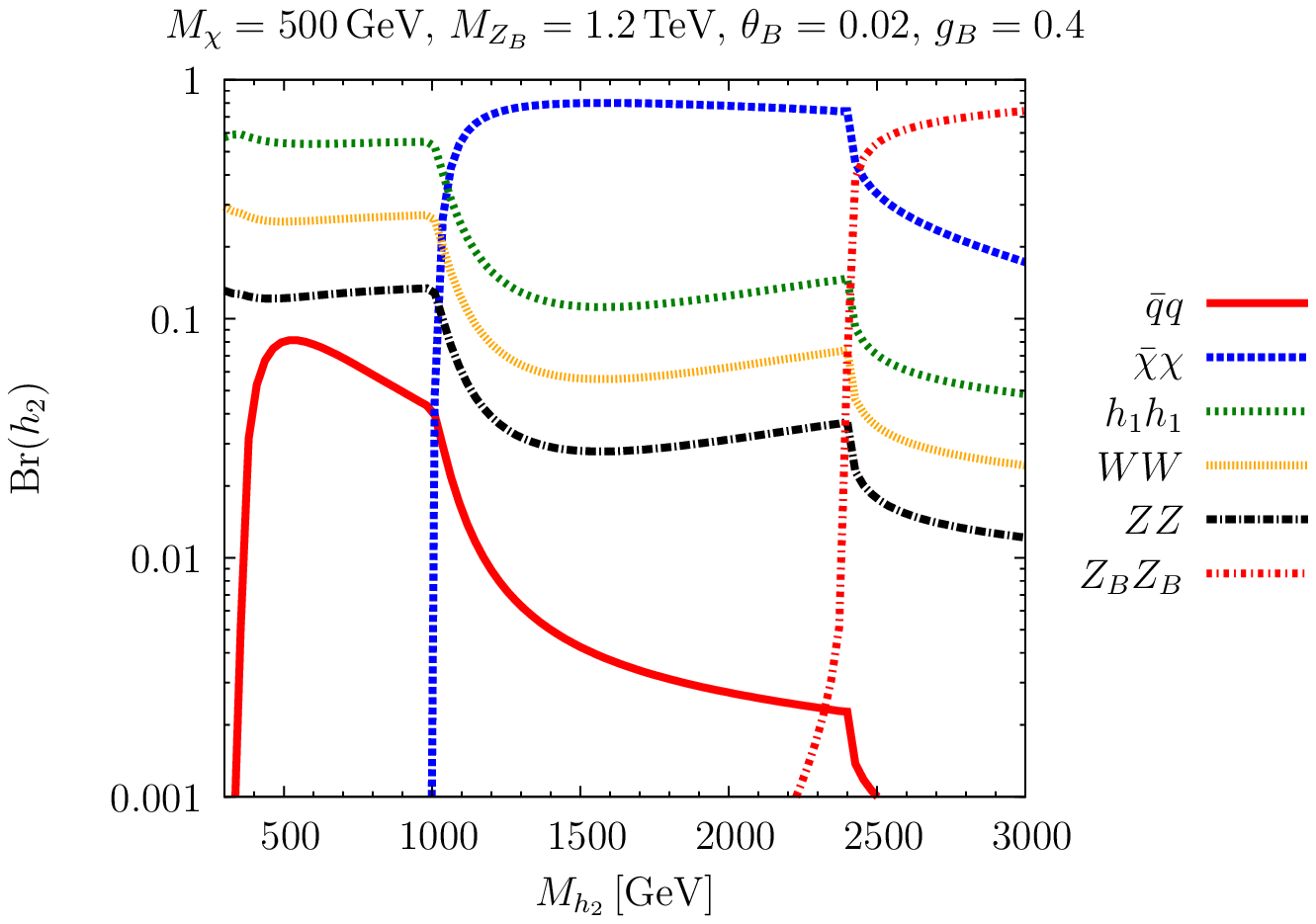}
 \caption{Decays of the heavy Higgs boson $h_2$. The left panel shows the total decay width  $\Gamma(h_2)$ of the $h_2$, and the right panel shows the different branching ratios. We use $M_\chi = \unit[500]{GeV}$, $M_{Z_B} = \unit[1.2]{TeV}$, $\theta_B = 0.02$, and $g_B = 0.4$ as parameters for the plots. Three-body decays are taken into account for the decay into two $Z_B$'s, $h_2 \to Z_B^\ast Z_B \to \bar{q} q Z_B$.}
 \label{fig:h2props}
\end{figure}

In Fig.~\ref{fig:h2props}, we show the properties of the decay of the heavy Higgs in more detail for $M_\chi = \unit[500]{GeV}$, $M_{Z_B} = \unit[1.2]{TeV}$, and $g_B = 0.4$. The choice of the small value of $\theta_B = 0.02$ is motivated by the above discussion. In the left panel, we show the total decay width as a function of the mass $M_{h_2}$, and in the right panel we display the branching ratios. Before the threshold for the decay into dark matter the decay into the SM Higgs dominates. As they become allowed, the decays into dark matter become dominant.  

The only viable production channel that is not suppressed is the associated $Z_B h_2$ production,
$$pp \  \to \  Z_B^* \  \to \ Z_B  h_2.$$
The $h_2$ can be produced through gluon fusion but this channel is suppressed by the mixing angle, like all other SM-like production channels. For example, when $\theta_B \approx 10^{-2}$ the gluon fusion production cross section will be suppressed by four orders of magnitude. In this case, in order to look for the decays into dark matter that we will be interested in, one needs to use a mono-jet or mono-photon, and the cross section for this process will be even more suppressed. The production of $h_2$ through the vector-boson fusion with the $Z_B$ gauge boson is not suppressed by the mixing angle but unfortunately it is suppressed by the $Z_B$ mass.
Therefore, we focus on the study of the associated production which is not suppressed by the mixing angle.

\begin{figure}
\includegraphics[width=0.49\linewidth]{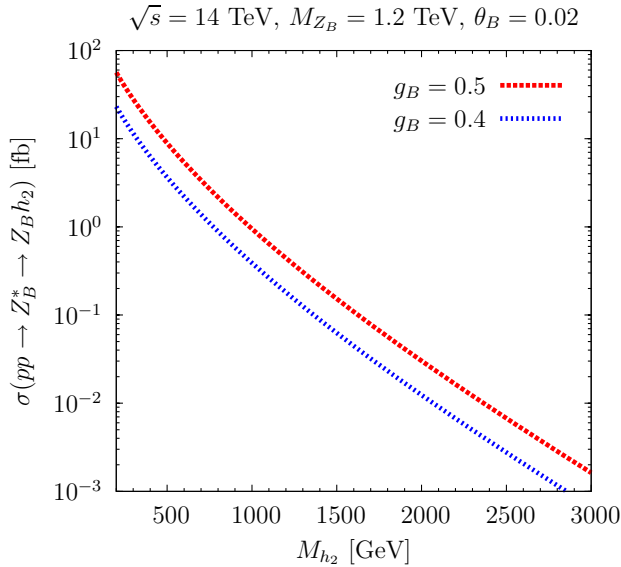}
 \caption{Associated production of the heavy Higgs $h_2$ at the LHC. We show the prediction for a center-of-mass energy of $\sqrt{s}=\unit[14]{TeV}$, and use $M_{Z_B} = \unit[1.2]{TeV}$, $\theta_B = 0.02$. The red line is for $g_B = 0.5$ and the blue line for $g_B = 0.4$.}
 \label{fig:associatedh2ZB}
\end{figure}

Using the cross section at the partonic level
\begin{align}
\sigma (\bar{q} q \to Z_B^* \to  Z_B h_2 ) (\hat{s}) &= \frac{g_B^4  \cos^2 \theta_B}{144 \pi \hat{s}^2} \frac{\left[  \hat{s}^2 - 2 \hat{s} (M_{Z_B}^2 + M_{h_2}^2) + (M_{Z_B}^2-M_{h_2}^2)^2 \right]^{1/2}}{\left[ (\hat{s} - M_{Z_B}^2)^2 + M_{Z_B}^2 \Gamma_{Z_B}^2 \right]} \nonumber \\
&\quad \times \left[ \hat{s}^2 + 2 \hat{s} (5 M_{Z_B}^2 - M_{h_2}^2) + (M_{Z_B}^2-M_{h_2}^2)^2 \right],
\end{align}
and the parton luminosities [see Eq.~\eqref{eq:partonluminosities}] we show in Fig.~\ref{fig:associatedh2ZB} the numerical results for this cross section. Only for large values of the gauge coupling $g_B$ can a significant production cross section be achieved. Notice that only $g_B = 0.4$ is allowed over the whole parameter space, see Fig.~\ref{fig:ZBbounds} for the experimental bounds.

\begin{figure}
\includegraphics[width=0.5\linewidth]{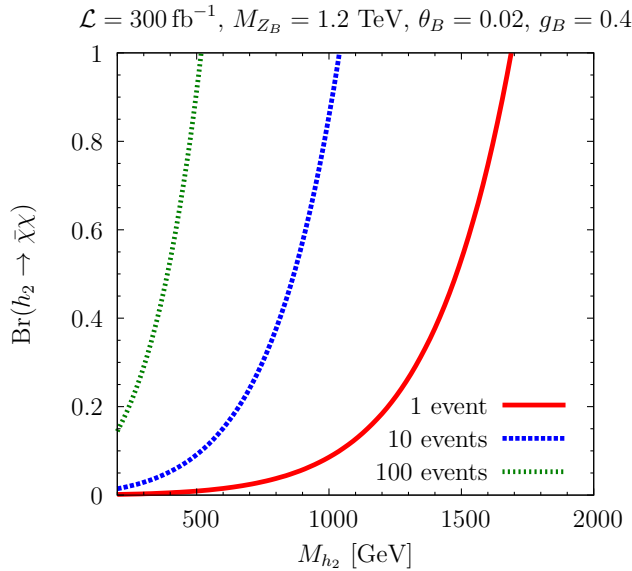}
\caption{Invisible decays of the heavy Higgs $h_2$. The plot shows the invisible branching ratio Br$(h_2 \to \bar{\chi} \chi)$ as a function of $M_{h_2}$ for different numbers of events $\bar{t} t E_\text{T}^\text{miss}$ (1 event in red, 10 events in blue, 100 events in green) at the LHC with a center-of-mass energy of $\sqrt{s} = \unit[14]{TeV}$ and a luminosity of $\mathcal{L}= \unit[300]{fb}^{-1}$. We use $M_{Z_B} = \unit[1.2]{TeV}$, $\theta_B = 0.02$, and $g_B = 0.4$. We show $M_{h_2}$ in the range from \unit[200]{GeV} to \unit[2]{TeV}.}
\label{fig:invBR}
\end{figure}

In this case one can have interesting signatures at the LHC with a $\bar{t} t$ pair and missing energy when $Z_B$ decays into two tops and $h_2$ decays into dark matter. The number of events for this channel is given by
\begin{equation}
N(\bar{t} t E_\text{T}^\text{miss})={\cal{L}} \times \sigma (pp \to Z_B h_2) \times \text{Br} (Z_B \to \bar{t} t) \times \text{Br}(h_2 \to \bar{\chi} \chi).
\end{equation}
In Fig.~\ref{fig:invBR}, we give Br$(h_2 \to \bar{\chi} \chi)$ for different numbers of expected events in $\bar{t} t E_\text{T}^\text{miss}$. Notice that we can say that naively the mass of the $h_2$ Higgs mass cannot be much beyond \unit[1]{TeV} in order to have a significant number of events. 
Using ${\cal{L}}=300 \ \rm{fb}^{-1}$, $M_\chi = \unit[500]{GeV}$, $M_{Z_B} = \unit[1.2]{TeV}$, $M_{h_2}= \unit[1.1]{TeV}$, $g_B = 0.4$, $\theta_B = 0.02$, and $B=1/2$, one has $ \sigma (pp \to Z_B h_2) = \unit[0.181]{fb}$, $\text{Br} (Z_B \to \bar{t} t)=0.137$, and $\text{Br}(h_2 \to \bar{\chi} \chi) = 0.452$. One therefore expects to have $N(\bar{t} t E_\text{T}^\text{miss}) = 3$ events for this channel and the given set of parameters. 
Still, for lower masses, the LHC could probe large fraction of the parameter space in this model by looking for missing energy and a $\bar{t} t$ pair.
Notice that the current collider bounds on $\bar{t} t E_\text{T}^\text{miss}$ are relevant for channels with QCD production cross sections, but not for our model. 

As discussed before, a key element to identify these events at the LHC is the reconstruction of the $Z_B$ gauge boson. In this way one can establish that the top quarks are 
from the $Z_B$ decays and the QCD background can be suppressed. In summary, the discovery of the leptophobic gauge boson $Z_B$, 
the Higgs $h_2$ with large branching ratio into dark matter, and the events with missing energy will be crucial to identify this model in the near future. 
\section{Baryonic Dark Matter}
\label{sec:dm}
In this theory one postulates the existence of a new sector needed to define an anomaly-free theory. One of the predictions is that in this new sector
the lightest field with fractional baryon number can describe the cold dark matter in the universe.  After the spontaneous breaking of local baryon number a remnant discrete ${\mathcal{Z}}_2$ symmetry 
protects the dark matter candidate $\chi$, which is a Dirac fermion. In this section we discuss the main properties of this dark matter candidate.
\subsection{Baryon Asymmetry vs.\ Dark Matter}
In models where we have the spontaneous breaking of local baryon number at the low scale, we must understand how it is possible to generate a baryon asymmetry in agreement with experiment. This issue has been investigated in great detail in Ref.~\cite{Hiren}, where the authors studied the solution for the chemical potentials in this theory. 
The analysis contains three key elements:
\begin{itemize}
\item The sphaleron condition on the chemical potentials is different because the sphaleron processes conserve total baryon number,
\begin{equation}
(Q_L Q_L Q_L \ell_L)^3 \overline{\Psi}_R \Psi_L, 
\end{equation}
which imposes the following condition on the chemical potentials: 
\begin{equation}
3 \left( 3 \mu_{u_L} + \mu_{e_L}\right) + \mu_{\Psi_L} - \mu_{\Psi_R}=0.
\end{equation}
\item There are two conserved global symmetries which can be used to protect the asymmetries. In the 
SM sector we have the $B-L$ symmetry and in the new sector we have the $\eta$ accidental global symmetry.
The new fermionic fields transform as 
\begin{eqnarray}
\Psi_{L,R} & \to & e^{i \eta} \  \Psi_{L,R}, \\
\eta_{L,R} & \to & e^{i \eta} \  \eta_{L,R}, \\
\chi_{L,R} & \to & e^{i \eta} \  \chi_{L,R},
\end{eqnarray}
under the $\eta$ symmetry.
\item Assuming chemical equilibrium and using all the interactions in the model, 
one can show~\cite{Hiren} that the relation between the baryon and the dark matter asymmetries is given by
\begin{equation}
\frac{n_q - n_{\bar{q}}}{\rm{s}}= r_1 \ \Delta (B-L)_\text{SM} + r_2 \ \Delta \eta,
\end{equation}
where
\begin{eqnarray}
r_1&=&\frac{32}{99}, \text{ and } r_2=\frac{15-14 B_2}{198}.
\end{eqnarray}
In the above equation $\Delta (B-L)_\text{SM}$ is the $B-L$ asymmetry generated through a mechanism such as leptogenesis, and 
\begin{equation}
\Delta \eta=\frac{n_\chi - n_{\bar{\chi}}}{\rm{s}},
\end{equation}
is the dark matter asymmetry. 
\end{itemize}

In general we can discuss three main scenarios to understand the relation between the baryon asymmetry and the dark matter relic density:
\begin{itemize}
\item $\Delta \eta=0$: In this case one has only a symmetric dark matter component and the baryon asymmetry is defined by the $B-L$ asymmetry,
\begin{equation}
\Omega_B = \frac{32}{99} \ \Omega_{B-L},
\end{equation}
where $\Omega_{B-L}={\rm{s}} \Delta(B-L)_\text{SM} M_p / \rho_c$ is normalized by the proton mass $M_p$. Here ${\rm{s}}$ is the entropy density and $\rho_c$ is the critical density. Notice that the coefficient is smaller than what one obtains for only the SM fields~\cite{Harvey:1990qw},
\begin{equation}
 \Omega_B^\text{SM} = \frac{28}{79} \ \Omega_{B-L}.
\end{equation}

The dark matter relic density in this case is only thermal; we will discuss the details in Section~\ref{sec:symmetricdm}.
\item $\Delta (B-L)_\text{SM}=0$: When there is no $B-L$ asymmetry in the SM sector, there is a simple relation between the baryon asymmetry 
and the dark matter asymmetry. In this case one needs to postulate a mechanism to generate the asymmetry in the dark matter sector. 
Imposing the condition $\Omega_{\chi} \leq 5 \Omega_\text{DM}$ one finds an upper bound on the dark matter mass:
\begin{equation}
M_\chi \leq \frac{5 (15 - 14 B_2)}{198} \  M_{p}. 
\end{equation}
\item In the general case when both asymmetries, $\Delta (B-L)_\text{SM}$ and $\Delta \eta$, are different from zero  one finds~\cite{Hiren} 
the following upper bound on the dark matter mass:
\begin{equation}
M_\chi \leq \frac{r_2 \  \Omega_\text{DM} \  M_p}{|\Omega_B -  r_1 \ \Omega_{B-L} |}.
\end{equation}
As discussed in Ref.~\cite{Hiren}, this scenario can be in agreement with the cosmological constraints. 
\end{itemize}

We have discussed the possible cases where one has a relation between the different asymmetries in this model. However, 
since in this context we do not have a simple mechanism to explain an asymmetry in the dark matter sector, we 
stick to the case $\Delta \eta=0$. Then, there is a simple connection between the $B-L$ asymmetry, 
generated by a mechanism such as leptogenesis, and the baryon asymmetry. In this case we can explain the 
observed cold dark matter relic density using standard thermal production. This is the main goal of the next section.
%
\subsection{Cold Symmetric Dark Matter}
\label{sec:symmetricdm}
Our dark matter candidate $\chi$ can annihilate into all the SM particles, as well as into the new Higgs $h_2$ and the leptophobic gauge boson $Z_B$. Therefore, we can 
have the annihilation channels
\begin{displaymath}
\bar{\chi} \chi \ \to \ \bar{q} q, \ \bar{\ell} \ell,\  WW, \ ZZ, \ h_i h_j, \ Z_B Z_B, \ h_i Z_B,
\end{displaymath}
where $h_i=h_1, h_2$. There are three main regimes for our study:
\begin{itemize}

\item $M_\chi < M_{Z_B}, M_{h_2}$: In this case the allowed annihilation channels into two 
SM fermions or gauge bosons are through the $Z_B$ gauge boson and the Higgses
\begin{displaymath}
\bar{\chi}  \chi \ \to \ Z_B^* \  \to \ \bar{q} q,
\end{displaymath} 
\begin{displaymath}
\bar{\chi}  \chi \ \to \ h^*_i \  \to \ \bar{q} q, \ \bar{\ell} \ell, \ WW, \ ZZ,
\end{displaymath} 
and into two SM Higgs bosons
\begin{displaymath}
\bar{\chi}  \chi \ \to \ h_1  h_1.
\end{displaymath} 
All the channels through the Higgs bosons and the annihilation into two SM Higgs bosons are velocity-suppressed, in addition to a suppression by the mixing angle.
Therefore, in most of the parameter space the annihilation through $Z_B$ will define the annihilation cross section 
allowed by the relic density constraints.

\item $M_{h_2} < M_{\chi} < M_{Z_B}$: In this scenario there are two extra allowed channels
\begin{displaymath}
\bar{\chi}  \chi \ \to \ h_1  h_2, \ h_2  h_2,
\end{displaymath} 
which are velocity-suppressed.
\item $M_{h_2}, M_{Z_B} < M_{\chi}$: Finally, if the dark matter is heavier than the $Z_B$ and $h_2$ 
bosons one can have new open channels which are not velocity-suppressed,
\begin{displaymath}
\bar{\chi} \chi \ \to \ h_i  Z_B, \ Z_B Z_B.
\end{displaymath} 
\end{itemize}

We have discussed in the previous section that in order to test this model one needs to discover the $Z_B$ gauge boson and the $h_2$ Higgs at the LHC. 
The decays $Z_B \to \bar{\chi} \chi$ and $h_2 \to \bar{\chi} \chi$ are crucial to identify this model since the dark matter candidate is present in the theory 
to cancel the baryonic anomalies. Therefore, this scenario has very important implications for the dark matter because when the decays 
$Z_B \to \bar{\chi} \chi$ and $h_2 \to \bar{\chi} \chi$ are allowed, the most important annihilation channel is $\bar{\chi} \chi \ \to \ Z_B^* \  \to \ \bar{q} q$.
The annihilation cross section for this channel is given by
\begin{multline}
\sigma (\bar{\chi} \chi \to  Z_B^*   \to  \bar{q} q) = \frac{3}{16 \pi s} \frac{\sqrt{s - 4 M_q^2}}{\sqrt{s - 4 M_\chi^2}} \\
\times \left\{ C_1^2 \left[ s^2 + \frac{1}{3} \left(s - 4 M_q^2\right) \left(s - 4 M_\chi^2\right) + 4 M_q^2 \left(s - 2 M_\chi^2\right) \right] 
+ 4 M_\chi^2 C_2^2 \left( s + 2 M_q^2 \right) \right\},
\end{multline}
where the coefficients $C_i$ are listed in Appendix~\ref{app:dm}.

The relic density can be computed using an analytic approximation~\cite{Kolb:1990vq,RD1,RD2},
\begin{equation}
\label{eq:relicdensity}
\Omega_\text{DM} h^2 = \frac{\unit[1.07 \times 10^{9}]{GeV}^{-1}}{J(x_f) \sqrt{g_\ast} \ M_\text{Pl}},
\end{equation}
where $M_\text{Pl}=\unit[1.22 \times 10^{19}]{GeV}$ is the Planck scale, $g_\ast$ is the total number of effective relativistic degrees of freedom at the time of freeze-out, and $J(x_f)$ is given by
\begin{equation}
J(x_f)=\int_{x_f}^{\infty} \frac{ \langle \sigma v \rangle (x)}{x^2} dx.
\end{equation}
The quantity $\langle \sigma v \rangle$ is a function of $x$, where $x=M_\chi/T$, and is given by
\begin{equation}
 \langle\sigma v\rangle (x) = \frac{x}{16 M_\chi^5 K_2^2(x)} \int_{4 M_\chi^2}^\infty \sigma \times ( s - 4 M_\chi^2) \ \sqrt{s} \ K_1 \left(\frac{x \sqrt{s}}{M_\chi}\right) ds.
\end{equation}
Notice that there is an additional factor 1/2 compared to the expression for $\langle\sigma v\rangle$ that is usually given, because we include particles and antiparticles, see the discussion in Ref.~\cite{RD1}. Therefore, the expression for the relic DM density in Eq.~\eqref{eq:relicdensity} describes the total DM relic density
\begin{equation}
 \Omega_\text{DM} = \Omega_\chi + \Omega_{\bar{\chi}}.
\end{equation}
The freeze-out parameter $x_f$ can be computed using
\begin{equation}
x_f= \ln \left( \frac{0.038 \ g \ M_\text{Pl} \ M_\chi \ \langle\sigma v\rangle (x_f) }{\sqrt{g_\ast x_f}} \right),
\end{equation}
where $g$ is the number of degrees of freedom. The modified Bessel functions $K_1(x)$ and $K_2 (x)$ are given by
\begin{eqnarray}
K_1(z)&=& z \int_{1}^{\infty} dt \ e^{-zt} (t^2-1)^{1/2}, \\
 K_2(z)&=& \frac{z^2}{3} \int_{1}^{\infty} dt \ e^{-zt} (t^2-1)^{3/2},
\end{eqnarray}
when ${\rm{Re}}(z) > 0$. 

The direct detection constraints must be included in order to understand which are the allowed values of the input parameters in this theory. 
The elastic spin-independent nucleon--dark matter cross section is given by
\begin{equation}
\sigma_{\chi N}^\text{SI} = \frac{M_N^2 M_\chi^2}{4 \pi (M_N + M_\chi)^2} \frac{g_B^4}{M_{Z_B}^4} B^2,
\end{equation}
where $M_N$ is the nucleon mass. Notice that $\sigma_{\chi N}^\text{SI}$ is independent of the matrix elements, because baryon number is a conserved current in the theory. The above equation can be rewritten as 
\begin{equation}
 \sigma_{\chi N}^\text{SI} (\text{cm}^2) = 3.1\times 10^{-41} \left( \frac{\mu}{\unit[1]{GeV}}\right)^2 \left( \frac{\unit[1]{TeV}}{r_B}\right)^4 B^2 \ \text{cm}^2,
\end{equation}
where $\mu = M_N M_\chi / (M_N + M_\chi)$ is the reduced mass and $r_B = M_{Z_B}/g_B$.

\begin{figure}[t]
\includegraphics[width=0.49\linewidth]{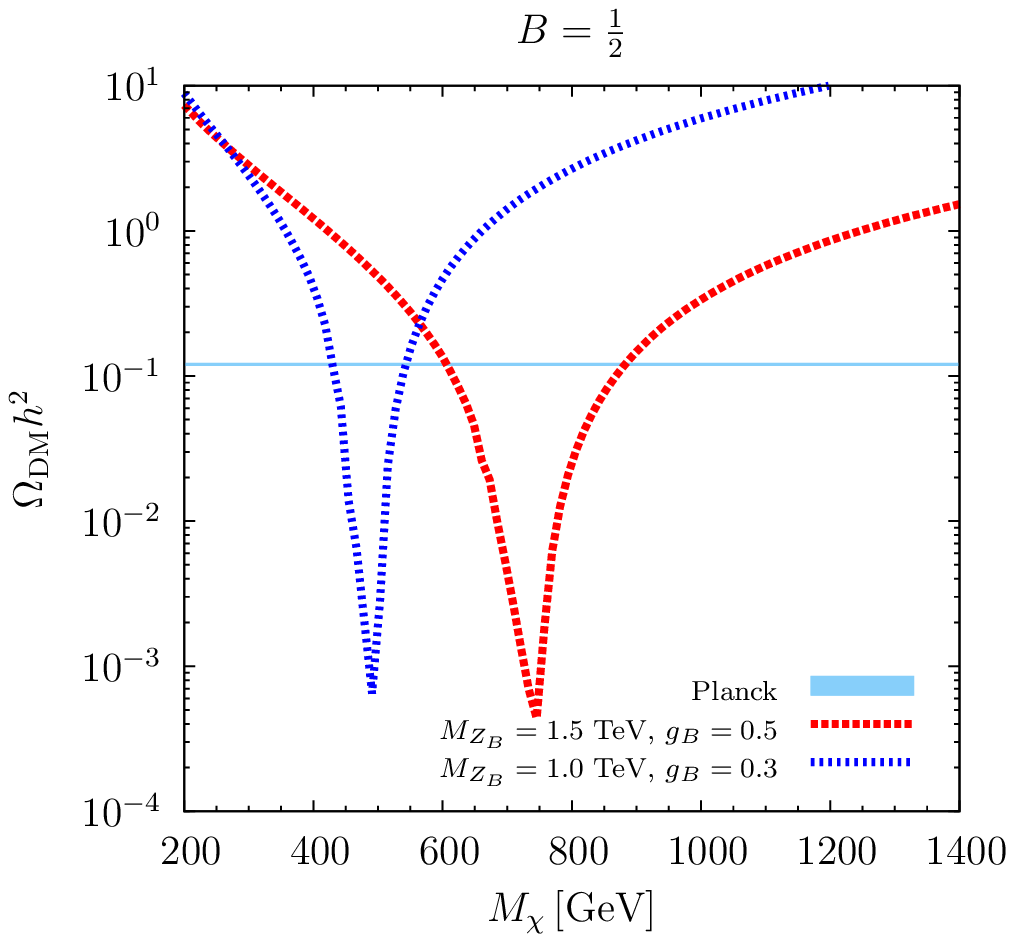}
\includegraphics[width=0.49\linewidth]{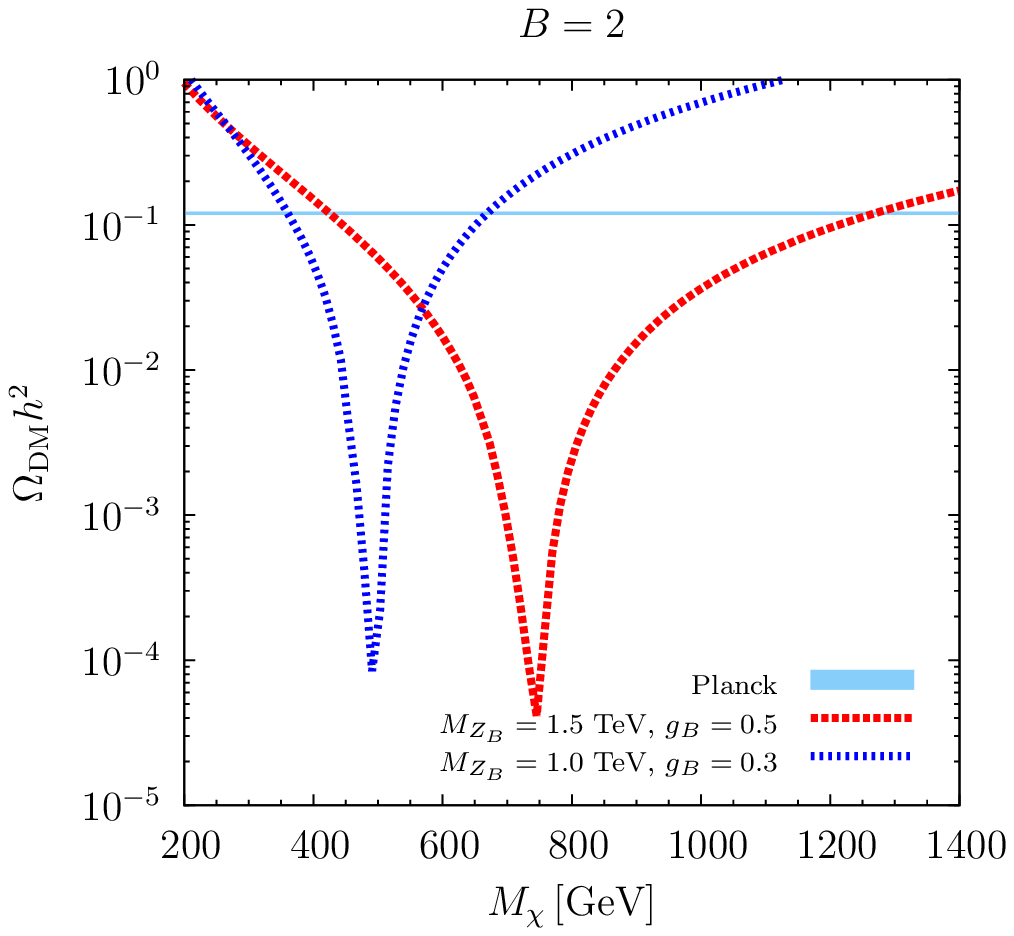}
\caption{Constraints from the DM relic density. The plots show the DM relic density $\Omega_\text{DM} h^2$ as a function of the DM mass $M_\chi$ for $M_{Z_B}=\unit[1.5]{TeV}$ and $g_B = 0.5$ (red) and for $M_{Z_B}=\unit[1.0]{TeV}$ and $g_B = 0.3$ (blue). The left panel is for $B = 1/2$, and the right panel is for $B=2$. The currently allowed value of $\Omega_\text{DM} h^2 = 0.1199 \pm 0.0024$~\cite{Ade:2013zuv} is marked by a thin blue band. }
\label{fig:relicdensity}
\end{figure}

\begin{figure}
\includegraphics[width=0.55\linewidth]{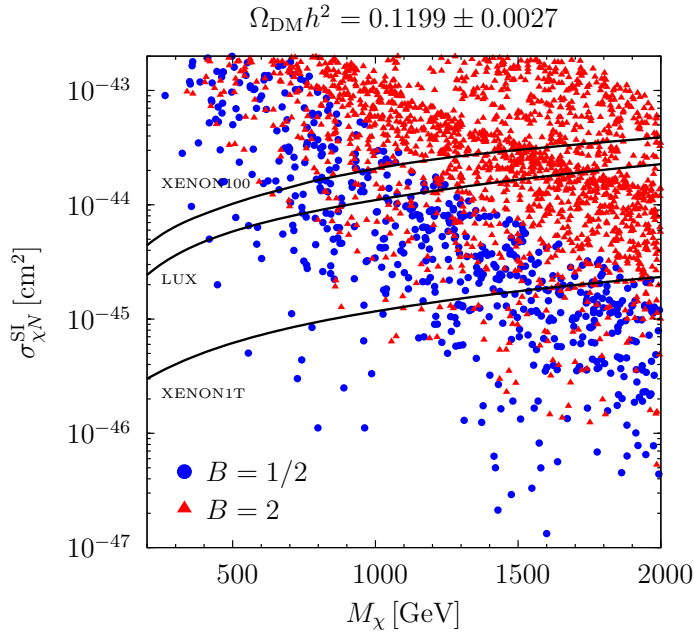}
\caption{Prospects for DM direct detection, assuming the value of the DM relic density $\Omega_\text{DM} h^2 = 0.1199 \pm 0.0027$ measured by Planck~\cite{Ade:2013zuv}. The plot shows the spin-independent elastic DM--nucleon cross section~$\sigma_{\chi N}^\text{SI}$ as a function of the DM mass $M_\chi$. The exclusion limits of XENON100~\cite{Aprile:2012nq} and LUX~\cite{Akerib:2013tjd} are given, as well as the projected limit for XENON1T~\cite{Aprile:2012zx}. The gauge coupling is varied inside $g_B \in [0.1,0.5]$, and the gauge boson mass is varied inside $M_{Z_B} \in [\unit[0.5]{TeV},\unit[5.0]{TeV}]$. Blue dots are for $B=1/2$, red triangles are for $B=2$.}
\label{fig:directdetection}
\end{figure}

In Fig.~\ref{fig:relicdensity} we show the relic density as a function of the DM mass $M_\chi$ for different choices of $g_B$ and $M_{Z_B}$. One can appreciate that one does not have to rely on the resonance to allow for the current value of the relic density measured by Planck~\cite{Ade:2013zuv}, $\Omega_\text{DM} h^2 = 0.1199 \pm 0.0024$. This result is important in order to show that even the naive estimation for the relic density gives good results in agreement with the cosmological constraints. 
In the case shown in the right panel when $B=2$ one finds solutions very far from the resonance because obviously the annihilation cross section is much larger than in the case when $B=1/2$. 

In Fig.~\ref{fig:directdetection} we show the values of the elastic spin-independent DM--nucleon cross section as a function of the DM mass, assuming that our dark matter candidate makes up the whole dark matter relic density. We take into account the constraints from XENON100~\cite{Aprile:2012nq} and LUX~\cite{Akerib:2013tjd}. Notice that the LUX bounds rule out many possible solutions. However, for $M_\chi \geq \unit[500]{GeV}$ one finds many solutions with the right dark matter relic density $\Omega_\text{DM}$ and in agreement with direct detection. As one can appreciate there are many viable solutions in agreement with the dark matter and collider experiments which can be used to understand the predictions for indirect detection. We will investigate the indirect signatures in a future publication.
%
\subsection{Upper Bound on the Symmetry Breaking Scale}
The existence of a non-zero relic density can be used to find an upper bound on the symmetry breaking scale, and we will discuss in detail how one can find this bound in what follows. Such a bound tells us that there is a possibility to test or rule out 
this model at current or future collider experiments. 

Neglecting the velocity-suppressed terms as well as the quark masses in the annihilation cross section we find that
\begin{equation}
\sum_q \sigma v (\bar{\chi} \chi \to Z_B^* \to \bar{q} q)  = \frac{g_B^4 M_{\chi}^2}{2 \pi} \frac{(B_1 + B_2)^2}{\left[ (4 M_\chi^2 - M_{Z_B}^2)^2+ M_{Z_B}^2 \Gamma_{Z_B}^2 \right]}.
\end{equation}
Using the upper bound on the dark matter relic density, $\Omega_\text{DM} h^2 \leq 0.12$, and the fact that the dark matter and the gauge boson masses are generated through the same mechanism, i.e., $M_\chi = \lambda_\chi v_B/\sqrt{2}$ and $M_{Z_B}=3 g_B v_B$, we find 
\begin{equation}
v_B^2 \leq \frac{g_B^4 \ \lambda_\chi^2 \ (B_1+B_2)^2 \ \unit[1.77\times10^9]{GeV}^2 }{\pi \ \left[ (2 \lambda_\chi^2 - 9 g_B^2)^2 + \frac{9}{4 \pi^2} g_B^8 \right] x_f}
\end{equation}
for a given value of $x_f$. Using this equation, it is possible to find an upper bound on the gauge boson mass which is given by
\begin{equation}
M_{Z_B} \leq 316.1 \  \frac{(B_1+B_2)}{\sqrt{x_f}} \ \unit{TeV}.
\end{equation}
However, typically $x_f$ takes values between $20$ and $40$.
Now, using $x_f=20$ and $B_1+B_2=1/2$ as an example, the upper bound on the gauge boson mass reads as
\begin{equation}
M_{Z_B} \leq \unit[35.3]{TeV},
\end{equation} 
and $M_\chi \leq \unit[17.7]{TeV}$.
Notice that this bound is much smaller than the one coming from unitarity~\cite{Griest:1989wd}. Therefore, we can say that this model could be tested in the near future.
\section{Summary}
\label{sec:summary}
We have investigated the main features of a simple theory where the baryon number is defined as a local gauge symmetry broken at the low scale.
This theory predicts the existence of a new gauge boson associated with baryon number which decays into the Standard Model quarks and dark matter.
We have shown the properties of the leptophobic gauge boson and the new Higgs boson decays. In both cases the branching ratio into dark matter 
can be large giving rise to signatures with missing energy at colliders. We have discussed the associated production $pp \to h_2 Z_B$ which is not suppressed by the mixing 
angle in the Higgs sector. Then, using the predictions of $h_2$ and $Z_B$ decays into dark matter and top quarks respectively, we have shown the possibility to 
test this theory at the LHC.

In order to find an upper bound on the leptophobic gauge boson mass we have discussed the bounds from the relic density constraints. 
We found that $M_{Z_B} \leq \unit[35.3]{TeV}$ when the freeze-out temperature is $x_f = 20$ and $B=1/2$. This bound is not very sensitive to the value of the freeze-out temperature 
and implies that we can rule out this theory in the near future at colliders.
We have shown the properties of all dark matter annihilation channels and the correlation with the collider constraints. 
The possibility to have a consistent scenario for baryogenesis has been analyzed. As one can see, combining the signatures at the LHC and 
dark matter constraints one could test this theory in current or future experiments. 

\section*{Acknowledgments}
P.F.P.\ thanks C.\ Cheung and M.\ B.\ Wise for discussions and the Caltech theory group for support and hospitality. We thank S.\ Ohmer and H.\ H.\ Patel for discussions.
\appendix
\section{Decay Widths}
\label{app:decay}

The partial decay widths of the leptophobic gauge boson $Z_B$ are given by
 \begin{align}
  \Gamma ( Z_B \rightarrow \bar{q} q ) &= \frac{g_B^2}{36 \pi} M_{Z_B} \left(1 - \frac{4 M_q^2}{M_{Z_B}^2} \right)^{\frac{1}{2}} \left(1 + \frac{2 M_q^2}{M_{Z_B}^2} \right), \\
   \Gamma ( Z_B \rightarrow \bar{\chi} \chi ) &= \frac{g_B^2 M_{Z_B}}{24 \pi} \left(1 - \frac{4 M_\chi^2}{M_{Z_B}^2} \right)^{\frac{1}{2}}  \left[ \left(B_1^2 + B_2^2\right) \left( 1- \frac{M_\chi^2}{ M_{Z_B}^2}\right) + 6 B_1 B_2 \frac{M_\chi^2}{ M_{Z_B}^2}\right].
 \end{align}

The partial decay widths of the new scalar boson $h_2$ read as
\begin{align}
\Gamma (h_2 \to \bar{f} f) &=\frac{N_f}{8 \pi} |c_{h_2 \bar{f} f}|^2 M_{h_2} \left( 1 \ - \ 4 \frac{M_f^2}{M_{h_2}^2} \right)^{3/2}, \\
\Gamma (h_2 \to \bar{\chi} \chi) &=\frac{9}{8 \pi} \frac{g_B^2 M_\chi^2}{M^2_{Z_B}} \cos^2 \theta_B M_{h_2} \left( 1 \ - \ 4 \frac{M_\chi^2}{M_{h_2}^2} \right)^{3/2}, \\
\Gamma (h_2 \to h_1 h_1) &= \frac{1}{32 \pi} \frac{|c_{112}|^2}{ M_{h_2}} \left( 1 - 4 \frac{M_{h_1}^2}{M_{h_2}^2} \right)^{1/2}, \\
\Gamma (h_2 \to W W) &= \frac{G_F}{8 \sqrt{2}\pi}  \sin^2 \theta_B \ M_{h_2}^3 \left( 1 \ - \ 4 \frac{M_W^2}{M_{h_2}^2} + 12 \frac{M_W^4}{M_{h_2}^4}\right) \left( 1 - 4 \frac{M_W^2}{M_{h_2}^2}\right)^{1/2} , \\
\Gamma (h_2 \to Z Z) &= \frac{G_F}{16 \sqrt{2}\pi}  \sin^2 \theta_B \ M_{h_2}^3 \left( 1 \ - \ 4 \frac{M_Z^2}{M_{h_2}^2} + 12 \frac{M_Z^4}{M_{h_2}^4}\right) \left( 1 - 4 \frac{M_Z^2}{M_{h_2}^2}\right)^{1/2} , \\
\Gamma (h_2 \to Z_B Z_B) &= \frac{1}{32 \pi}  \frac{\cos^2 \theta_B \ M_{h_2}^3}{v_B^2} \left( 1 \ - \ 4 \frac{M_{Z_B}^2}{M_{h_2}^2} + 12 \frac{M_{Z_B}^4}{M_{h_2}^4}\right) \left( 1 - 4 \frac{M_{Z_B}^2}{M_{h_2}^2}\right)^{1/2}.
\end{align}
\section{Production Mechanisms at the LHC}
\label{app:production}
The average amplitudes integrated over solid angle for the processes $\bar{q}q \to Z_B^\ast \to Z_B h_2$ and $\bar{q}q \to Z_B^\ast \to \bar{t} t$ are given by
\begin{align}
 \int d \Omega \ |\bar{\mathcal{M}} (\bar{q} q \to Z_B h_2 )|^2 &= \frac{4 \pi g_B^4 \cos^2 \theta_B}{9} \, \frac{\left[s^2 + 2 s (5 M_{Z_B}^2 - M_{h_2}^2) + (M_{Z_B}^2 - M_{h_2}^2)^2 \right]}{\left[ (s-M_{Z_B}^2)^2 + M_{Z_B}^2 \Gamma_{Z_B}^2\right]},  \\
 \int d \Omega \ |\bar{\mathcal{M}} (\bar{q} q \to Z_B^* \to  \bar{t} t )|^2 &= \frac{16 g_B^4 \pi s}{243} \frac{\left[  2 M_t^2 + s \right]}{\left[ (s - M_{Z_B}^2)^2 + M_{Z_B}^2 \Gamma_{Z_B}^2 \right]}.
\end{align}
%
\section{Dark Matter Annihilation Channels}
\label{app:dm}
The annihilation of our dark matter candidate $\chi$ into two SM quarks is mediated by the s-channel exchange of the leptophobic gauge boson and the two physical Higgs particles,
\begin{equation*}
\bar{\chi}  \chi \  \to \ Z_B^*, h_i^* \ \to \ \bar{q} q, 
\end{equation*}
where $h_i=h_1,h_2$. The average amplitude squared for these channels is given by (a color factor $N_c = 3$ for the quarks is taken into account)
\begin{align}
\int d \Omega \ |\bar{\mathcal{M}} (\bar{\chi} \chi \to \bar{q} q )|^2 &= 12 \pi \ C_1^2 \left( s^2 + \frac{1}{3} (s - 4 M_q^2) (s - 4 M_\chi^2) + 4 M_q^2 (s - 2 M_\chi^2) \right) \nonumber \\
&\quad + 48 \pi M_\chi^2 \ C^2_2 \ (s + 2 M_q^2) + 12 \pi |C_3|^2 (s - 4 M_q^2) (s-4 M_\chi^2), 
\end{align}
where
\begin{align}
C_1^2 &= \frac{g_B^4}{18} \frac{(B_1^2 + B_2^2)}{[(s-M_{Z_B}^2)^2 + M_{Z_B}^2 \Gamma_{Z_B}^2]}, \\
C_2^2 &= \frac{g_B^4}{9} \frac{B_1 B_2}{[(s-M_{Z_B}^2)^2 + M_{Z_B}^2 \Gamma_{Z_B}^2]}, \\
C_3&= i \frac{c_{h_1 \bar{\chi} \chi} c_{h_1 \bar{q} q}}{s - M_{h_1}^2+i M_{h_1} \Gamma_{h_1}} + i \frac{c_{h_2 \bar{\chi} \chi} c_{h_2 \bar{q} q}}{s - M_{h_2}^2+i M_{h_2} \Gamma_{h_2}}. 
\end{align} 
The couplings are given by
\begin{align}
 c_{h_1 \bar{\chi} \chi} &= 3 g_B \frac{M_\chi}{M_{Z_B}} \sin\theta_B,\\
 c_{h_2 \bar{\chi} \chi} &= 3 g_B \frac{M_\chi}{M_{Z_B}} \cos\theta_B,\\
 c_{h_1 \bar{q} q} &= \frac{M_q}{v_0} \cos\theta_B,\\
 c_{h_2 \bar{q} q} &=\frac{M_q}{v_0} \sin\theta_B .
\end{align}

\section{Kinetic Mixing}
\label{app:kineticmixing}
In this appendix we calculate the kinetic mixing generated once the heavy fields are integrated out. The fields $\Psi$ and $\eta$ are charged 
under $U(1)_Y$ and $U(1)_B$, and once they are integrated out a kinetic mixing is generated at one-loop level. The relevant interactions for our discussion are 
\begin{equation}
 \mathcal{L} \supset - g_1 \bar{f} \gamma^\mu \left( Y_L P_L + Y_R P_R \right) f B_\mu -  g_B  \bar{f} \gamma^\mu \left( B_L P_L \ + \ B_R P_R \right)  f \  Z_\mu^B, 
\end{equation}
where $f=\eta, \Psi$, $g_1$ is the $U(1)_Y$ gauge coupling and $B_\mu$ is the corresponding gauge field. The hypercharges $Y_L$ and $Y_R$ as well as the baryon numbers $B_L$ and $B_R$ for the lepto-baryons can be read off from Table~\ref{tab:ModelI}. Including all the lepto-baryons we obtain the following expression for the kinetic mixing:
\begin{equation}
i \Pi_{Y,B}^{\mu\nu} = i \frac{g_1 g_B}{24 \pi^2} \sum_f   C_f \left[2 A_0(m_f) - (q^2 + 2m_f^2) B_0(q,m_f,m_f) + \frac{1}{3} q^2 -2 m_f^2 \right] \left(g^{\mu\nu} - \frac{q^\mu q^\nu}{q^2}  \right),
\end{equation}
where
\begin{align}
 C_\Psi &= -(B_1 + B_2),\ \ \
 C_\eta = -\frac{1}{2} (B_1 + B_2).
\end{align} 
The functions $A_0$ and $B_0$ are the usual scalar Passarino--Veltman integrals.

In the limit $q^2 \rightarrow 0$, the kinetic mixing reads as
\begin{equation}
 \epsilon_B^\text{1-loop} =  \frac{1}{12 \pi^2} g_B g_1\sum_f C_f \ln \left( \frac{\mu_\text{R}}{m_f}\right),
\end{equation}
where $\mu_\text{R}$ is the renormalization scale. As we will see the lepto-baryons with masses beyond the TeV only give a small contribution to the kinetic mixing. 
Using $M_\Psi = \unit[10]{TeV}$, $\mu_\text{R} = \unit[500]{GeV}$, $B=1/2$, $g_1=0.35$ and $g_B=0.4$, we obtain
\begin{equation}
 \epsilon_B=0.002
\end{equation}
for the contribution of $\Psi$. The contribution of the $\eta$ field is even smaller if one uses the same mass. 
As one can appreciate the kinetic mixing is very small in this model. Therefore, the phenomenological analysis presented 
in this article is justified in this way.


\end{document}